\documentclass[
    prd, twocolumn, superscriptaddress, nofootinbib, amsmath, amssymb,
    aps, floatfix,preprintnumbers
]{revtex4-2}

\usepackage[utf8]{inputenc}
\usepackage{graphicx,xcolor,setspace}
\usepackage[
    colorlinks=true,
    linkcolor=blue,
    urlcolor=blue,
    citecolor=blue
]{hyperref}

\usepackage{amsfonts}
\usepackage{physics}
\usepackage{array}
\newcommand{\PreserveBackslash}[1]{\let\temp=\\#1\let\\=\temp}
\newcolumntype{C}[1]{>{\PreserveBackslash\centering}p{#1}}
\newcolumntype{R}[1]{>{\PreserveBackslash\raggedleft}p{#1}}
\newcolumntype{L}[1]{>{\PreserveBackslash\raggedright}p{#1}}
\usepackage{changes}

\begin{document}

\title{Fully \textit{ab-initio} all-electron calculation of dark matter--electron scattering in crystals with evaluation of systematic uncertainties}
\date{\today}

\author{Cyrus E. Dreyer}
\affiliation{Department of Physics \& Astronomy, Stony Brook University, Stony Brook, NY 11794, USA}
\affiliation{Center for Computational Quantum Physics, Flatiron Institute, 162 Fifth Avenue, New York, NY 10010, USA}

\author{Rouven Essig}
\affiliation{C. N. Yang Institute for Theoretical Physics, Stony Brook University, Stony Brook, NY 11794, USA}

\author{Marivi Fernandez-Serra}
\affiliation{Department of Physics \& Astronomy, Stony Brook University, Stony Brook, NY 11794, USA}
\affiliation{Institute for Advanced Computational Sciences, Stony Brook University, Stony Brook, NY 11794, USA}

\author{Aman Singal}
\affiliation{Department of Physics \& Astronomy, Stony Brook University, Stony Brook, NY 11794, USA}
\affiliation{C. N. Yang Institute for Theoretical Physics, Stony Brook University, Stony Brook, NY 11794, USA}
\affiliation{Institute for Advanced Computational Sciences, Stony Brook University, Stony Brook, NY 11794, USA}

\author{Cheng Zhen}
\affiliation{Department of Physics \& Astronomy, Stony Brook University, Stony Brook, NY 11794, USA}
\affiliation{C. N. Yang Institute for Theoretical Physics, Stony Brook University, Stony Brook, NY 11794, USA}

\preprint{YITP-SB-2023-11}

\begin{abstract}
    We calculate target-material responses for dark matter--electron scattering at the \textit{ab-initio} all-electron level using atom-centered gaussian basis sets.  The all-electron effects enhance the material response at high momentum transfers from dark matter to electrons, $q\gtrsim \order{10\ \alpha m_e}$, compared to calculations using conventional plane wave methods, including those used in \texttt{QEDark}; this enhances the expected event rates at energy transfers $E \gtrsim 10$~eV, especially when scattering through heavy mediators.  We carefully test a range of systematic uncertainties in the theory calculation, including those arising from the choice of basis set, exchange-correlation functional, number of unit cells in the Bloch sum, $\mathbf{k}$-mesh, and neglect of scatters with very high momentum transfers.  We provide state-of-the-art crystal form factors, focusing on silicon and germanium.  Our code and results are made publicly available as a new tool, called \textit{Quantum Chemistry Dark} (``\texttt{QCDark}''). 
\end{abstract}

\maketitle

\section{Introduction}

    There has been rapid progress in direct-detection searches of sub-GeV dark matter (DM) by looking for electron recoils from DM-electron scattering in noble liquids and crystals~(see, e.g.,~\cite{Essig:2011nj,Essig:2012yx,Essig:2017kqs,Angle:2011th,Essig:2015cda, Lee:2015qva,Crisler:2018gci,sensei2019, SENSEI2020, DAMIC2020, Agnese:2018col,Aguilar-Arevalo:2019wdi,Arnaud:2020svb,Agnes:2018oej,Aprile:2016wwo,Aprile:2019xxb,XENON:2021qze,PandaX-II:2021nsg,Blanco:2019lrf,DAMIC-M:2022aks, SuperCDMS2020,XENON1T2020,Tiffenberg:2017aac,Settimo:2018qcm,Castello-Mor:2020jhd,oscuraExperiment}). A theoretical description of such DM--electron scattering processes requires a quantitative description of the electronic structure of the detector material. This can be achieved by utilizing methods from condensed-matter physics and quantum chemistry. In particular, density density functional theory (DFT) has been demonstrated to be a powerful tool for determining the ground-state electronic structure in a wide variety of materials from first principles, as well their response to various perturbations~\citep{Martin2020}. Also, DFT is the necessary first step to performing calculations using more advanced methods to treat systems with, e.g., stronger electron-electron correlations~\cite{Martin2016}
    
    Several choices need to be made when calculating the electronic structure and DM--electron scattering rate from DFT.  This includes choosing the type of basis functions to describe the electronic wavefunctions; whether to separate the core electrons in the material from  the valence electrons by, e.g., pseudopotentials~\cite{Martin2020} or the projector-augmented wave (PAW) method~\cite{PAW}; and the exchange-correlation (XC) functional, which incorporates the many body effects of the electron--electron interactions~\cite{KohnSham65,Martin2020}. These choices often represent trade-offs between accuracy and computational efficiency~\cite{dick21}, which differ for different types of materials. Moreover, once these choices are made, the relevant properties must be tested for convergence with respect to the numerical parameters of the calculation.
    
    DFT-based techniques have been applied to calculating DM--electron scattering in a variety of materials relevant for detectors including semiconductors \citep{Essig:2015cda, Knapen2021Diel, Knapen2022DarkELF, Griffin21, Trickle2022}, semimetals~\cite{Hochberg:2015fth}, superconducting nanowire single-photon detectors~\cite{Hochberg:2019cyy,Hochberg:2021yud}, quantum dots~\cite{blanco2022dark}, etc. For the most part, studies of electron recoils in semiconductors have taken the approach of a plane wave basis set, core electrons frozen in pseudopotentials of PAW potentials, and either local/semilocal XC functionals based on the local density approximation (LDA), the generalized gradient approximation (GGA), or hybrid functionals that include a fraction of exact electron-electron exchange interaction~\cite{Kahn:2021ttr}. DFT using plane wave basis sets are the one most commonly used for studying solids in condensed-matter physics and materials science.
    
    However, there are motivations for choosing an alternative approach. First, it was recently shown that explicit treatment of the core electrons significantly affects the DM--electron scattering rates~\cite{Liang:2018bdb,Griffin21,Trickle2022}. While sub-GeV DM does not typically excite an electron from a core orbital to the conduction band, the inclusion of the rapidly oscillating part of the valence all-electron wavefunction near the atomic cores (where it must be orthogonal to the core electron wavefunctions ~\cite{HEINE19701}) is necessary to capture DM--electron scattering events with high momentum transfer. For a calculation with a plane wave basis to be tractable, these oscillations must be smoothed through the use of pseudopotentials, though the all-electron wavefunction can be reconstructed if PAWs are used~\cite{Griffin21,Trickle2022}.
    
    A basis set made up of localized functions, e.g., atom-centered Gaussians, can treat core and valence electrons on the same footing without significant increase in computational cost. In addition, such basis sets can be used either with periodic boundary conditions for solids~\cite{PySCF2020} or for finite systems, such as molecules or nanostructures~\cite{PySCF2020}, increasing the flexibility to explore different DM detector materials. Finally, using localized basis sets allows the use of quantum chemistry methods, which allows many-body correlations to be included when calculating the wave functions for atoms, molecules, liquids, and solids.
    
    In this work, we develop the computational methodology to perform all-electron calculations of DM--electron scattering based on localized Gaussian basis sets. The resulting code, which we call \textit{Quantum Chemistry Dark} (\texttt{QCDark}), is based on the python-based simulations of chemistry framework (PySCF)~\cite{PySCF2015, PySCF2018, PySCF2020} package, which allows for DFT and quantum chemistry methods to be used on both finite and extended systems .\footnote{The code is available at \href{https://github.com/asingal14/QCDark}{https://github.com/asingal14/QCDark}.} Via benchmark calculations on silicon and germanium, we show that the basis sets can be converged. PySCF has previously been employed in the context of DM--electron scattering in isolated atoms and molecules in \cite{Hamaide:2021hlp}. We compare our results for DM-electron scattering with previous work, and find good agreement with EXCEED-DM~\cite{Griffin21, Trickle2022}, which reconstructs all--electron effects with PAW reconstruction. We also quantify uncertainties related to the choice of XC functional and numerical convergence parameters. 
    
    The rest of the paper is organized as follows. In \S\ref{sec:2}, we describe electronic structure calculations, include a comparison between plane-wave and atomic-centered bases, discuss how to include all--electron effects, and describe the DM--electron scattering rate calculations with quantum chemistry basis sets. In \S\ref{sec:4}, we describe the results for Si and Ge, including the systematic uncertainties and the effects of the secondary ionization modeling; we also calculate the annual modulation rates, and compare our results with those of previous works.  We conclude in \S\ref{sec:6}. Two appendices contain technical information, including the properties of Cartesian Gaussians (Appendix~\ref{app:cart_gauss}) and a derivation of the scattering rate formulae (Appendix~\ref{app:derivation}). 
    
\section{Calculating Dark Matter-Electron Scattering Rates}\label{sec:2}
    
    In this section, we introduce the computational approach that underlies \texttt{QCDark}, comparing the all-electron localized basis set approach used in this work with previous implementations based on plane waves and pseudopotentials.
    
    \subsection{Electronic structure}\label{subsec:2.1}
        The electronic structure of the material is described by Kohn-Sham (KS)~\cite{KohnSham65} DFT, in which the equations that describe the system of many interacting electrons are mapped onto a set of single-particle equations in an effective potential constructed to reproduce the ground-state electron density and total energy. The energy functional of the density $n$ is given by (Gaussian units are assumed throughout) 
        \begin{equation}
            \begin{split}
                E[n] &= T_{\text{S}}[n] + \int\mathrm{d}\mathbf{r} v_{\text{ext}}(\textbf{r})n(\textbf{r}) \\ &+ 
                \frac{1}{2}\int\mathrm{d}\mathbf{r}\int\mathrm{d}\mathbf{r}^\prime\ \frac{n(\mathbf{r})n(\mathbf{r}^\prime)}{\abs{\mathbf{r - r}^\prime}} + E_{\text{xc}}[n] \ ,
                \label{eq:KS}
            \end{split}
        \end{equation}
        where $T_{\text{S}}$ is the sum of the kinetic energies of the non-interacting orbitals; $v_{\text{ext}}$ is the external potential given by, e.g., the atomic nuclei in an all-electron calculation or the ions if pseudopotentials are used; and $E_{\text{xc}}$ is the so-called exchange-correlation energy, which accounts for the many-body and quantum effects that are neglected in  $T_{\text{S}}$ and the Hartree electron-electron interaction (third term in Eq.~(\ref{eq:KS}))~\cite{Martin2020}. Writing the density as a sum over auxiliary single-particle orbitals, i.e., $n(\textbf{r})=\sum_i\vert\psi_i(\textbf{r})\vert^2$, and minimizing Eq.~(\ref{eq:KS}) with respect to variations in $\psi$ results in the KS equations, 
        \begin{equation}
            \left[-\frac12 \nabla^2+v_{\text{eff}}(\textbf{r})\right]\psi_i(\textbf{r})=\epsilon_i\psi_i(\textbf{r})\,,
            \label{eq:KS2}
        \end{equation}
        where $v_{\text{eff}}=v_{\text{ext}}+\int d\mathbf{r}^\prime \frac{n(\mathbf{r})n(\mathbf{r}^\prime)}{\abs{\mathbf{r -r}^\prime}}+\delta E_{\text{xc}}[n]/\delta n(\textbf{r})$, with the last term being the functional derivative of the exchange-correlation energy with respect to the density, which results in the XC potential $v_{\text{xc}}(\textbf{r})$. The KS equations in Eq.~(\ref{eq:KS2}) must be solved self-consistently, since the effective potential depends on the density.
        
        Several choices exist in the above calculation, which we discuss in subsequent sections. First, the KS wavefunctions $\psi$ must be expressed in terms of some basis functions (\S\ref{subsubsec:2.1.1}); the choice of these basis functions has a significant impact on many other aspects of the calculation, including the possible choices for the boundary conditions and the treatment of core electrons (\S\ref{subsubsec:2.1.2}). Also, the exact form for the exchange-correlation potential is not known, and the choice of approximate  $v_{\text{xc}}(\textbf{r})$ can result in qualitatively different results for the electronic structure of the material (\S\ref{subsubsec:2.1.3}). 
    
        \subsubsection{Basis sets}\label{subsubsec:2.1.1}
        Previous works~\cite{Essig:2015cda, Knapen2021Diel, Knapen2022DarkELF, Griffin21, Trickle2022} used plane waves as basis function (PW-basis), together with periodic boundary conditions. Then the KS wavefunctions for a given wavevector \textbf{k} in the first Brillouin Zone can be written as
        \begin{equation}
            \psi_{i\mathbf{k}} (\mathbf{r}) = \sqrt{\frac{1}{V}}\sum_{\mathbf{K}}u_i (\mathbf{k+K})e^{i\mathbf{(k+K)\cdot r}}\,,
        \end{equation}
        where $\textbf{K}$ is a reciprocal lattice vector, $e^{i\mathbf{(k+K)\cdot r}}$ are the basis functions, $u_i (\mathbf{k+K})$ are the coefficients, and $V$ is the volume of the crystal. The accuracy of the PW-basis for describing $\psi$ is governed by the number of reciprocal lattice vectors included in the basis, which is usually specified as a kinetic energy cutoff of the plane waves. The PW-basis is good at describing relatively delocalized states, e.g., the states near and above the Fermi level in most solids; however, as we will discuss in \S\ref{subsubsec:2.1.2}, it is computationally expensive to capture the core electrons with a PW basis, since their localized nature requires the inclusion of very high energy plane waves and hence a large PW-basis size. 
        
        In this work, we use atom-centered Cartesian Gaussian basis sets. These basis sets are efficient at treating localized states, including the core electrons, and may be used for periodic or finite systems. The building blocks of this basis are \emph{primitive} Gaussians,
        \begin{equation}\label{eq:cartestian_gauss}
            \begin{split}
                G_{ijk} (\mathbf{r}, \xi_\mu, \mathbf{A}) &= (x - A_x)^i (y - A_y)^j (z - A_z)^k \\
                &\times \exp[ - \xi_\mu (\mathbf{r} - \mathbf{A})^2]\,,
            \end{split}
        \end{equation}
        where \textbf{A} is the atomic position, $\xi_\mu$ is an adjustable parameter, and the Cartesian exponents $i$, $j$, and $k$ are all integers that satisfy the condition $i+j+k = l$, with $l$ being the angular quantum number of the shell. Then, the basis functions are given by \emph{contracted} Gaussians, i.e., weighted sums of $N_{\rm prim}$ primitive Gaussians:
        \begin{equation}
            \Tilde{G}_{\alpha} (\mathbf{r}) = \sum_{\mu = 1}^{N_{\rm prim}}N_\mu c_\mu G_{ijk}(\mathbf{r}, \xi_\mu, \mathbf{A})\,,
        \end{equation}
       where $\alpha=\{\kappa,ijk\}$ is a composite index that runs over all nuclei $\kappa$ in the system (or unit cells for periodic boundary conditions) as well as the Cartesian exponents, $N_\mu$ is the normalization of the primitive Gaussian, and $c_\mu$ is the coefficient of the primitive Gaussian. Note that $N_\mu$ and $c_\mu$ are fixed throughout the DFT calculation.
        
        For periodic calculations, the atomic orbitals in the unit cell are then
        \begin{equation}\label{eq:blochsum}
            \phi_{\alpha \mathbf{k}}\left(\mathbf{r}\right) = \sum_\mathbf{R} e^{i\mathbf{k}\cdot\mathbf{R}}\Tilde{G}_\alpha\left(\mathbf{r - R}\right),
        \end{equation}
        where \textbf{R} are the real-space lattice vectors. Upon self-consistently solving the DFT hamiltonian, we obtain a coefficient matrix for each $\mathbf{k}$, $C_{i\alpha} (\mathbf{k})$, so that the Kohn-Sham wavefunctions (often referred to as molecular orbitals) are
        \begin{equation}\label{eq:molecularorbitals}
            \psi_{i\mathbf{k}} (\mathbf{r}) = \frac{1}{\sqrt{N_\mathrm{cell}}}\sum_\alpha C_{i\alpha}(\mathbf{k}) \phi_{\alpha\mathbf{k}} (\mathbf{r}),
        \end{equation}
        where $N_\mathrm{cell}$ is the number of unit cells in the crystal.
        
        Compared to plane waves, Gaussian basis sets are significantly more complex. First of all, Gaussian basis sets are element-specific and the total basis set will be the sum of those from the individual atoms. For a given element, the construction of the basis set requires defining $c_\mu, N_\mu$, and $\xi_\mu$ for each atomic shell $n$ and angular momentum $l = i + j+ k$. The choice of basis presents a source of systematic uncertainty in our calculation, which we estimate by varying the basis sets for each system.
        
        The size of the basis set determines the number of orbitals, with multiple important parameters to consider. Moreover, there are several naming conventions that are widely used. In this paper, we use several basis sets, all taken from~\cite{BasisSetExchange}. The number of ``zetas'' in a basis set refers to the number of orbitals as a function of the number of occupied orbitals. An $N$-zeta (NZ) basis set would have $N$ basis functions for each fully or partly occupied orbital in an atomic species. Si has an electronic configuration of $\rm 1s^2 2s^2 2p^6 3s^2 3p^2$ (where the $\rm 3s^23p^2$ are valence electrons), and so has 3 s-orbitals and $2\times3$ p-orbitals, and so a DZ (double zeta) basis set would have 6 s- and $4\times3$ p-orbitals. The treatment of $l\geq2$ orbitals differs between spherical and cartesian gaussians. This is because there exists a linear combination of, for example, three d-orbitals in Cartesian Gaussians $(ijk) = (200),\ (020), \text{ and }(002)$ orbitals, which mimics (in this case) an s-orbital. In addition, certain basis sets contain $N$ zeta only for the valence orbitals, with one atomic orbital for core states. This is called a NZ-valence or NZV basis set. One can further add polarization functions on the valence orbitals, which adds new orbitals. For example, for Si, a DZP (double-zeta polarized) basis set would add a d-orbital to allow for the electrons to be polarized in the atoms.
        In general, addition of extra orbitals and polarization functions improves convergence, though diffuse components in basis sets may become pseudo-linearly dependent in periodic calculations~\cite{PySCF2018, PySCF2020}. 
        
        It should be noted that the aforementioned details represent a very brief overview of the very complex field of quantum chemical basis sets~\cite{Dunning89}, and different basis sets, especially with different naming conventions can add complexities to these considerations.

        We employ the TZP and the def2-TZVP basis set for Si and Ge respectively, and show the uncertainties associated with the choice of basis set in \S\ref{subsubsec:basis_uncertainty}.
        
        \subsubsection{Treatment of core electrons in DFT}\label{subsubsec:2.1.2}
        The treatment of core electrons, i.e., those tightly bound to the nuclei, is closely connected to the choice of basis set. For atom-centered Gaussians, it is straightforward to include all of the electrons in the DFT calculation, since localized basis functions can just as easily describe core electrons as those near and above the Fermi level. 
        However, for plane waves, describing such localized wavefunctions would require a prohibitively large energy cutoff. In addition, including core orbitals would require the wavefunctions of the valence electrons to be orthogonalized to the wavefunctions of the core electrons; this  would introduce rapid oscillations in the region around the atomic nuclei, which would also require too many plane waves to be computationally tractable. Therefore, plane-wave calculations usually freeze the core orbitals via pseudopotentials, effective core potentials (ECPs), or the projector-augmented wave (PAW) method. This has three main effects. First, the core orbitals do not participate in hybridization and bonding in the crystal; this is usually an excellent approximation, since such orbitals are so tightly localized around the atoms, so that there is negligible overlap between atoms. Second, since they are not explicitly included, DM--electron scattering transitions between core orbitals and the conduction band are neglected; this is also not usually an issue, since the energies of such transitions is often beyond the scope of light DM searches. Third, and most crucially, the ``pseudowavefunctions'' are smooth in the core region, since they no longer must be explicitly orthogonalized to the core orbitals. 
        
        As was shown in~\cite{Liang:2018bdb,Griffin21} (see also \S 4), all-electron effects are crucial for describing DM--electron scattering events with high momentum transfer, since such events couple high-frequency modes, which are only present in the all-electron valence and conduction bands due to the rapid oscillations near the core. In~\cite{Griffin21}, the all-electron wavefunctions were recovered after a plane-wave calculation via the PAWs. In \S\ref{sec:4}, we benchmark this methodology against the full all-electron calculation allowed by our Gaussian basis set. 
            
        \subsubsection{Exchange--correlation functional}\label{subsubsec:2.1.3}
        In practice, the exact form of the exchange and correlation energy in Eq.~(\ref{eq:KS}) is  not known; however, there are many well-motivated approximations (see~\cite{LibXC}). The choice of exchange and correlation functionals presents a major source of systematic uncertainty in the calculation. This systematic uncertainty can be estimated by calculating the electronic structure with different functionals, and comparing the results. The broad categories of functionals are
        \begin{enumerate}
            \item Local Density Approximation (LDA) -- $E_{xc}$ is a functional of only $n$, $E_{xc}\longrightarrow E_{xc}[n].$
            \item Generalized Gradient Approximation (GGA) -- $E_{xc}$ is a functional of $n$ and $\grad n$, $E_{xc}\longrightarrow E_{xc}[n, \grad n]$.
            \item Meta-GGAs (mGGA) contain higher order derivatives of $n$, including terms like $\partial\cdot\partial n$ and $\laplacian n$. 
            
            \item Hybrid functional are GGA and mGGA functionals with added exact Hartree-Fock exchange.
            \item Double hybrid functionals add M\o ller-Plesset perturbation theory at second order (``MP2 level'') to hybrid functionals in an effort to better model correlations. 
        \end{enumerate}
        
        We use the well tested PBE0 hybrid exchange-correlation functional. We further test several GGAs, mGGAs, and other hybrid functionals to show the dependence of DM--electron scattering rates on the choice of exchange-correlation functionals. Moreover, the choice of $E_\mathrm{xc}$ affects the calculated bandgap $E_\mathrm{gap}$ of the material, and so we apply a scissor correction to match the experimental bandgap. 
        
    \subsection{Excitation rates in atom-centered bases}\label{subsec:2.2}

        \subsubsection{Theory}\label{subsec:2.2.1}
        As described in \S \ref{subsec:2.1}, we self-consistently solve the Kohn-Sham equations to obtain the coefficients $C_{i\alpha}(\mathbf{k})$ in Eq.~\eqref{eq:molecularorbitals}. The key quantity required from the DFT calculation for calculating DM--electron scattering is the crystal form factor (equivalent to Eq.~(3.17) of~\cite{Essig:2015cda}), 
        \begin{widetext}
            \begin{equation}\label{eq:theoryFF}
                \begin{split}
                    \left|f_{\rm crystal}(q, E_e)\right|^2 = \frac{2\pi^2}{E_e}\frac{1}{\alpha m_e^2 V_\mathrm{cell}}\sum_{ij}\int_{\rm BZ}\frac{V_\mathrm{cell}d^3k}{(2\pi)^3}&\frac{V_\mathrm{cell}d^3k^\prime}{(2\pi)^3}E_e\delta\left(E_e - (E_{j\mathbf{k}^\prime} - E_{i\mathbf{k}})\right) \times\\
                    &\left.\sum_\mathbf{K^\prime}q\delta\left(q -\left|\mathbf{k^\prime+K^\prime - k}\right|\right)\left|f_{\left[j\mathbf{k}^\prime,i\mathbf{k}\right]}(\mathbf{q})\right|^2\right|_{\theta_q = \theta_U, \phi_q = \phi_U}\,,
                \end{split}
            \end{equation}
        \end{widetext}
        where $E_{i\mathbf{k}}$ is the energy of the $i^\mathrm{th}$ orbital from ground state at $\mathbf{k}$ in the reciprocal cell and $\mathbf{U = k^\prime +K^\prime - k}$. Here, $E_e$ and $q$ are the energy and momentum transferred from the DM particle to the electron, and $\theta_v$ and $\phi_v$ refer to the polar and azimuthal angles of $\mathbf{v}$, respectively. Indices $i$ and $j$ run over occupied and unoccupied orbitals, respectively, and 
        \begin{equation}\label{eq:mat_elems}
            \begin{split}
                f_{\left[j\mathbf{k}^\prime,i\mathbf{k}\right]}(\mathbf{q}) = \sum_\mathbf{R}e^{-i\mathbf{k}^\prime\cdot\mathbf{R}} \int d^3 r\ & C^\dagger_{j\beta}(\mathbf{k}^\prime)\Tilde{G}^*_\beta(\mathbf{r - R}) \\\times&e^{i\mathbf{q\cdot r}}\ \Tilde{G}_\alpha (\mathbf{r})C_{\alpha i}(\mathbf{k})\, .
            \end{split}
        \end{equation}
        The DM--electron scattering rate for a DM particle of mass $m_\chi$ and local density $\rho_\chi$ is
        \begin{equation}\label{eq:finRate}
            \begin{split}
                \frac{d R_\mathrm{crystal}}{d \ln{E_e}} = \frac{\rho_\chi}{m_\chi}& N_\mathrm{cell}\Bar{\sigma}_e\alpha\frac{m_e^2}{\mu_{\chi e}^2}\int d\ln{q} \frac{E_e}{q} \eta\left(v_\mathrm{min}(q, E_e)\right)\\&\times\abs{F_\chi (q)}^2\abs{f_\mathrm{crystal}(q, E_e)}^2 \abs{f_e/f_e^0}^2,
            \end{split}
        \end{equation}
        where
        \begin{equation*}
            \sigma_e (q)= \Bar{\sigma}_e \abs{F_\chi(q)}^2 = \Bar{\sigma}_e \left(\frac{(\alpha m_e)^2 + m_V^2}{q^2 + m_V^2}\right)^2
        \end{equation*}
        is the DM--electron interaction cross section assuming a bosonic mediator with mass $m_V$. In general, we assume two limits, $m_V\gg q$ (called the heavy-mediator limit) and $m_V\ll q$ (called the light-mediator limit). The factor $\abs{f_e/f_e^0}^2$ is a screening factor discussed next (see also \cite{Griffin21}), while $\eta(v_\mathrm{min}(q, E_e))$ is the average inverse speed of DM in the galaxy (for more details, see~\cite{Essig:2015cda} or Appendix~\ref{app:derivation}) and $\mu_{\chi e}^{-1} = m_\chi^{-1} + m_e^{-1}$. We describe the calculation of the matrix elements, Eq.~\eqref{eq:mat_elems}, in Appendix~\ref{app:cart_gauss}. 

        If the DM--electron interaction is mediated by a dark photon or scalar, the interaction is screened due to the in-medium effects~\cite{Hochberg:2015fth, Gelmini:2020xir}. Recent works \citep{Knapen2021Diel,Hochberg2021Diel} have emphasized the importance of this electrostatic screening, especially for recoils at low energy transfer. In the results shown below, we follow the prescription of~\cite{Griffin21, Trickle2022}, and multiply the crystal form factor, $\abs{f_\mathrm{crystal}}^2$ by a factor of $\abs{f_e/f_e^0}^2$, with $f_e/f_e^0 = \left(\mathbf{\hat{q}\cdot\epsilon\cdot\hat{q}}\right)^{-1}$. Here $\mathbf{\epsilon}$ is the dielectric function, which is modelled as~\cite{screening}
        \begin{equation}\label{eq:screening}
            \begin{split}
                \epsilon (q, E_e)&\ = 1 + \\&\left[\frac{1}{\epsilon_0 - 1} + \tau \left(\frac{q}{q_\mathrm{TF}}\right)^2 + \frac{q^4}{4m_e^2\omega_p^2} - \left(\frac{E_e}{\omega_p}\right)^2\right]^{-1}\,,
            \end{split}
        \end{equation}
        where $\epsilon_0 \equiv \epsilon(0,0)$ is the empirically measured static dielectric constant, $\tau$ is a fitting parameter (we use the results from~\cite{screening}, which fit their dielectric function to the results of~\cite{Walter:1970dielectric}), $\omega_p$ is the plasma frequency, and $q_\mathrm{TF}$ is the Thomas-Fermi momentum, with values listed in Table~\ref{tab:screening}. This equation ignores the tensorial nature of $\epsilon$, since the crystals we consider here are cubic. This equation only estimates the real part of the dielectric function, and we do not account for the imaginary part in our screening. 
        
        Ref.~\cite{Trickle2022} compares the differences in the expected DM--electron scattering rates using the analytical model in Eq.~\eqref{eq:screening} for the dielectric function with an RPA calculation of the dielectric function. We expect an $\order{10\%}$ correction for creating 3 or fewer electrons-hole pairs ($Q \leq 3e^-$) with a numerically calculated dielectric function using a preliminary RPA dielectric function calculation, and $\order{1\%}$ or less for creating more than 3 electron-hole pairs.

        \begin{table}[t!]
            \caption{Parameters used in the dielectric function calculation in Eq.~\eqref{eq:screening} for Si and Ge from~\cite{screening}.}
            \centering
            \begin{tabular}{l@{\hspace{1cm}}c@{\hspace{0.4cm}}c@{\hspace{0.4cm}}c@{\hspace{0.4cm}}c}\hline\hline
                Target&$\epsilon_0$&$\tau$&$\omega_p$ [eV]&$q_\mathrm{TF}$[KeV]\\\hline
                Si & 11.3 & 1.563 & 16.6 & 4.13 \\
                Ge & 14.0 & 1.563 & 15.2 & 3.99 \\\hline\hline
            \end{tabular}
            \label{tab:screening}
        \end{table}

        \subsubsection{Numerical implementation}\label{subsubsec:2.2.2}
        In practice, the numerical integration in $\mathbf{k}$-space requires the replacement
        \begin{equation}
            \int_{\mathrm{BZ}}\ \frac{V_\mathrm{cell}\ d^3k}{(2\pi)^3} \left(\dots\right) \longrightarrow  \frac{1}{N_\mathbf{k}}\sum_\mathbf{k} \left(\dots\right)\,,
        \end{equation}
        where $N_\mathbf{k}$ is the number of $\mathbf{k}-$points in the chosen $\mathbf{k}-$grid.
        
        Moreover, to discretize $\abs{f_\mathrm{crystal}(q, E_e)}^2$ in $q$ and $E_e$, we define bins with width $\Delta q$ and $\Delta E_e$ and bin centers $\left\{q_n\right\}$ and $\left\{E_e^m\right\}$, respectively. Then, the discretization procedure follows as
        \begin{equation}
            \begin{split}
                \abs{f_\mathrm{crystal} (q_n, E_e^m)}^2 &\equiv \int_{q_n - \frac{1}{2}\Delta q}^{q_n + \frac{1}{2}\Delta q}\frac{dq}{\Delta q}\times \\&\int_{E_e^m - \frac{1}{2}\Delta E_e}^{E_e^m + \frac{1}{2}\Delta E_e}\frac{dE_e}{\Delta E_e}\ \abs{f_\mathrm{crystal} (q, E_e)}^2.
            \end{split}
        \end{equation}
        Consequently, the numerical crystal form factor is
        \begin{widetext}
            \begin{equation}\label{eq:numericalFF}
                \begin{split}
                    \abs{f_\mathrm{crystal} (q_n, E_e^m)}^2 = \frac{2\pi^2}{E_e^m}\frac{1}{\alpha m_e^2 V_\mathrm{cell}}\frac{1}{N_\mathbf{k}^2}&\sum_{ij}\sum_{\mathbf{k,k^\prime}}\sum_{\mathbf{K^\prime}} \frac{E_e^m}{\Delta E_e}\frac{q_n}{\Delta q} \abs{f_{\left[j\mathbf{k}^\prime,i\mathbf{k}\right]}(\mathbf{k^\prime + K^\prime - k})}^2\times\\
                    & \Theta\left(1 - \frac{\abs{E_{j\mathbf{k^\prime}} - E_{i\mathbf{k}} - E_e^m}}{\frac{1}{2}\Delta E_e}\right) \Theta\left(1 - \frac{\abs{\abs{\mathbf{k^\prime + K^\prime - k}} - q_n}}{\frac{1}{2}\Delta q}\right).
                \end{split}
            \end{equation}
        \end{widetext}
        Here, as for Eq.~\eqref{eq:theoryFF}, the recoil electron is assumed to be transferred from occupied molecular orbital $\ket{i, \mathbf{k}}$ to unoccupied orbital $\ket{j, \mathbf{k^\prime + K^\prime}}$. 

        \begin{figure*}
            \centering
            \includegraphics[width=0.497\linewidth]{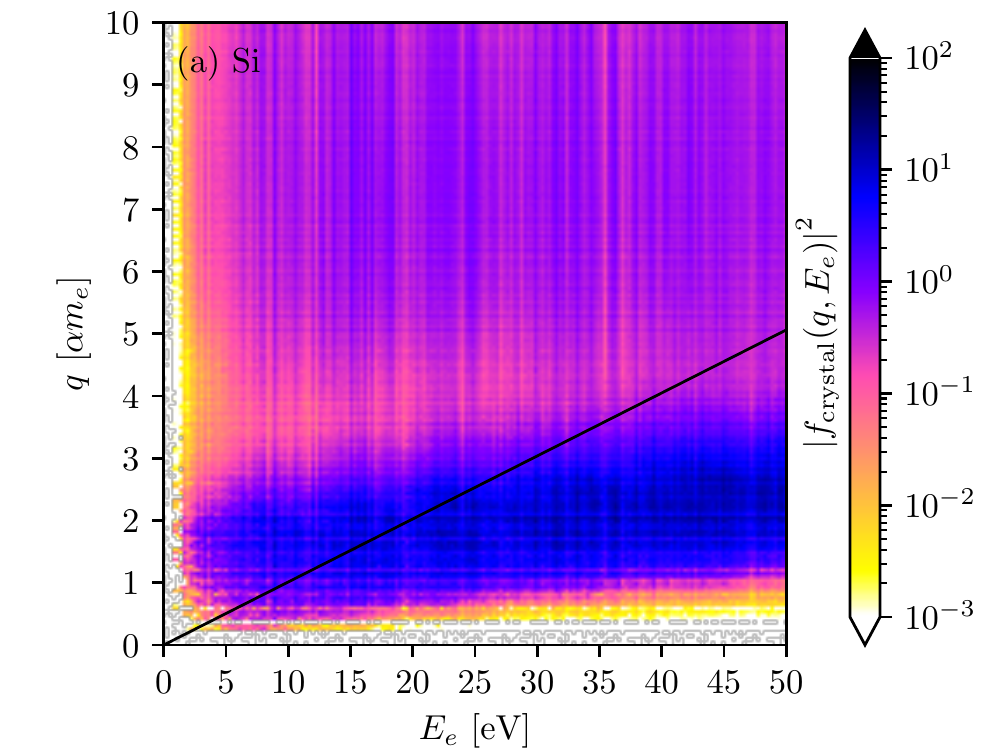}
            \includegraphics[width=0.497\linewidth]{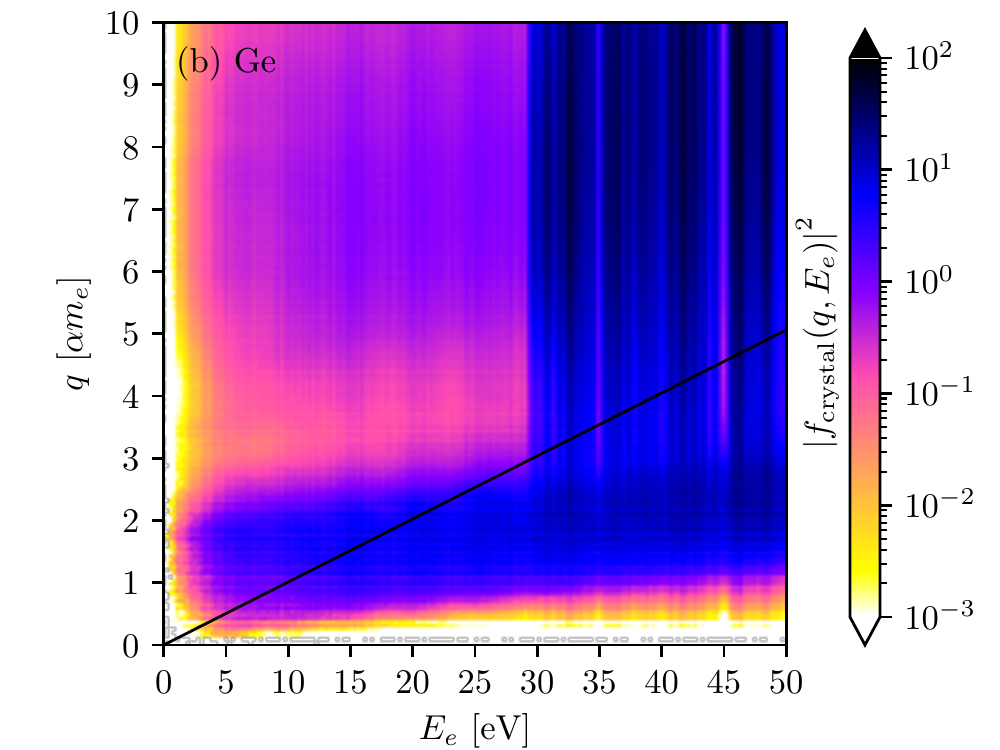}
            \caption{Panels (a) and (b) show the calculated crystal form factor $\abs{f_\mathrm{crystal} (q, E_e)}^2$ (see Eq.~\eqref{eq:numericalFF}) for silicon and germanium, respectively. The details of the calculation are described in \S\ref{subsec:4.1}. The region beneath the black line is kinematically inaccessible for halo DM, as it would require $v_\mathrm{min} (q, E_e) > v_\mathrm{Escape} + v_\mathrm{Earth}$ (see Appendix \ref{app:derivation} for more details).}
            \label{fig:cff}
        \end{figure*}

        In principle, there should be $n_k = N_T$ points modelled in the calculation, where $N_T$ is the number of unit cells in this crystal, with $N_T \gtrsim \order{10^{23}}$ for a 10 g crystal. However, such a dense mesh is impractical. Since we need to calculate rates for transition from each $\mathbf{k}$ to each $\mathbf{k}^\prime$, the computational complexity scales as $n_k^2$, and the calculation quickly becomes unfeasible.
        
        We generally model the grid in reciprocal space as a $\Gamma$-centered Monkhorst-Pack grid, with $n_{k,i}$ points along the $b_i$ reciprocal lattice vector. We use the shorthand $n_{k,1}\times n_{k,2}\times n_{k,3}$ to denote the mesh in the reciprocal space, which has $n_k = \prod_{i = 1}^3 n_{k,i}\ \mathbf{k}-$points. Because both Si and Ge crystallize in FCC cells, it is reasonable to set $n_{k,1} = n_{k,2} = n_{k,3}$. We choose $n_{k,i} = 4$ and $6$ for Si and Ge, respectively, but check the dependence of DM--electron scattering rates on our choice of $\mathbf{k}-$grid in \S\ref{subsubsec:k-grid_uncertainty}. 
        
        In addition, it becomes computationally expensive to include a large number of $\mathbf{K}$ vectors, and so in practice, we must limit the vectors to some $q_\mathrm{max}$. The choice of $q_\mathrm{max}$ is another source of systematic error. We choose $q_\mathrm{max} = 25 \alpha m_e$ and $q_\mathrm{max} = 20 \alpha m_e$ for Si and Ge, respectively, and further show the dependence of DM--electron scattering rates on $q_\mathrm{max}$ in \S\ref{subsubsec:qmax_uncertainty}. 
    
\section{Results \& discussion}\label{sec:4}

    In this section, we describe the results from our calculation of the crystal form factors (Eq.~\eqref{eq:numericalFF}) and the DM--electron scattering rates (Eq.~\eqref{eq:finRate}) for Si and Ge performed with \texttt{QCDark}. We evaluate the various systematic uncertainties, and compare our results to those from other available codes. Finally, we look at the annual modulation rate as a function of the DM mass, $m_\chi$ and the DM form factor, $F_\chi$.

     Table \ref{tab:dft_pars} lists the values of crystal parameters used for our calculation of DM--electron scattering rates in both Si and Ge, including the experimental bandgap used for the scissor correction procedure. 
    \begin{table}[b]
        \centering
        \caption{Parameters used for DFT calculation of electronic structure of Si and Ge crystals. $E_\mathrm{gap}$ refers to the scissor corrected bandgap we employ for the materials.}
        \begin{tabular}{l@{\hspace{1cm}}l@{\hspace{0.4cm}}c@{\hspace{0.4cm}}c}\hline\hline
            Target&Crystal&lattice&$E_\mathrm{gap}$ [eV]\\
            & structure&constant [\AA]&\\\hline
            Si & FCC & 5.43 & 1.11 \\
            Ge & FCC & 5.65 & 0.67 \\\hline\hline
        \end{tabular}
        \label{tab:dft_pars}
    \end{table}

    \subsection{Crystal Form Factor}\label{subsec:4.1}
        
        The calculated crystal form factors using Eq.~\eqref{eq:numericalFF} are shown in Fig.~\ref{fig:cff}, for silicon and germanium crystals in the left and right panels respectively. The region below the black lines are kinematically inaccessible for halo DM, i.e., the halo DM, irrespective of the mass of the fermion, is unable to transfer energy $E_e$ with a momentum transfer $q < E_e/(v_\mathrm{Escape}+v_\mathrm{Earth})$.
        
        Both panels show an enhancement of the crystal form factor at $q\gtrsim 4\ \alpha m_e$ compared to Fig.~5 of~\cite{Essig:2015cda}, which is a consequence of including all--electron effects in our calculation. The enhancement in the crystal form factor for germanium crystals at $E_e \gtrsim 30$~eV corresponds to the transitions from the semi-core 3d-shell to the conduction bands, which has an energy of $-28.6$~eV relative to the top of the valence bands in our calculation. 
        
    \subsection{Dark matter--electron scattering rates}
        \begin{figure*}
            \centering
            \includegraphics[width=0.497\linewidth]{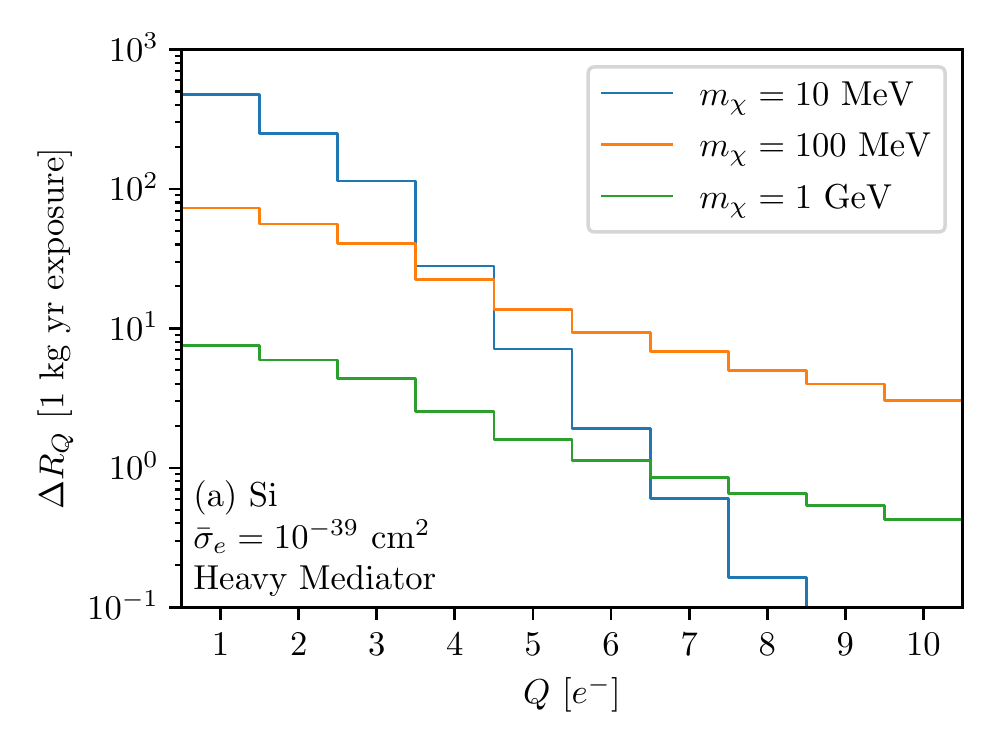}
            \includegraphics[width=0.497\linewidth]{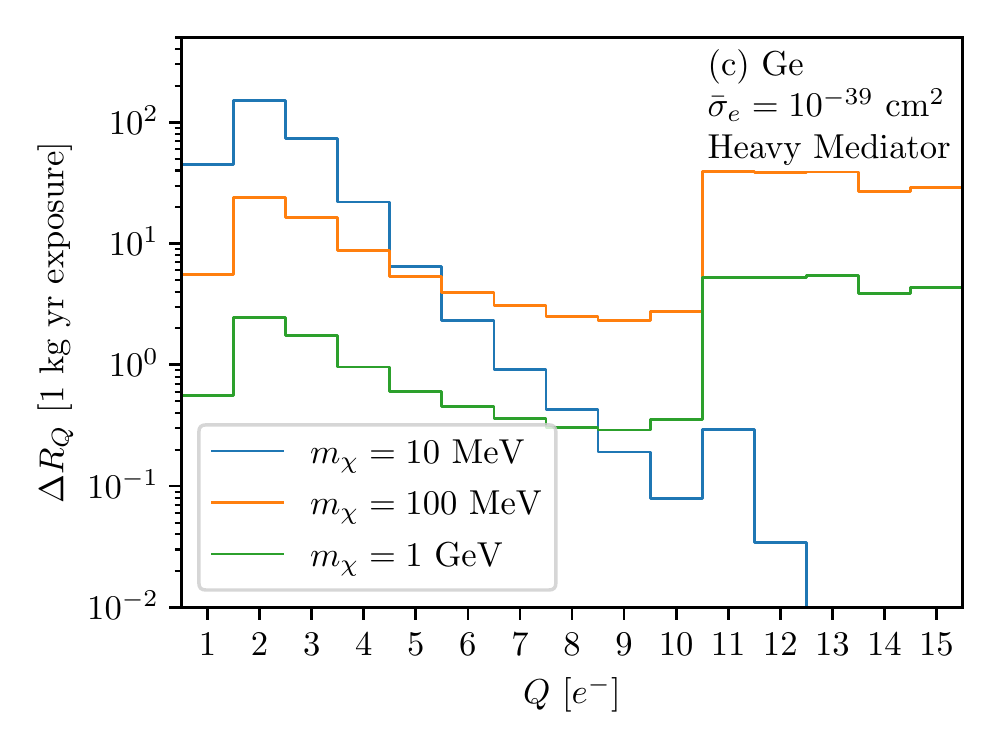}
            \includegraphics[width=0.497\linewidth]{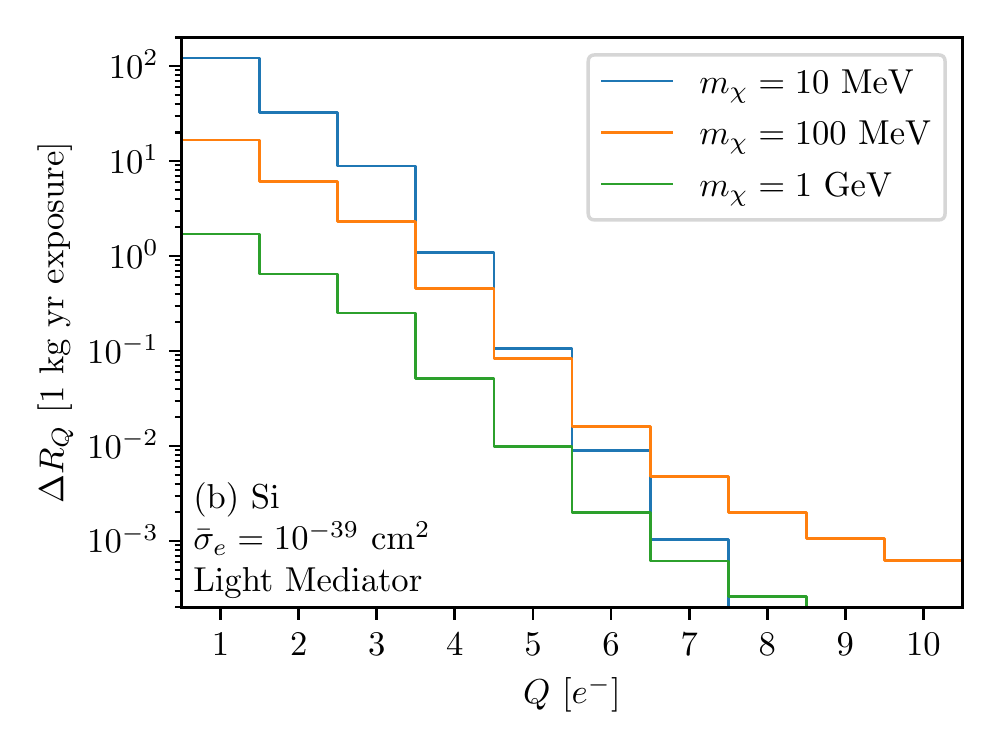}
            \includegraphics[width=0.497\linewidth]{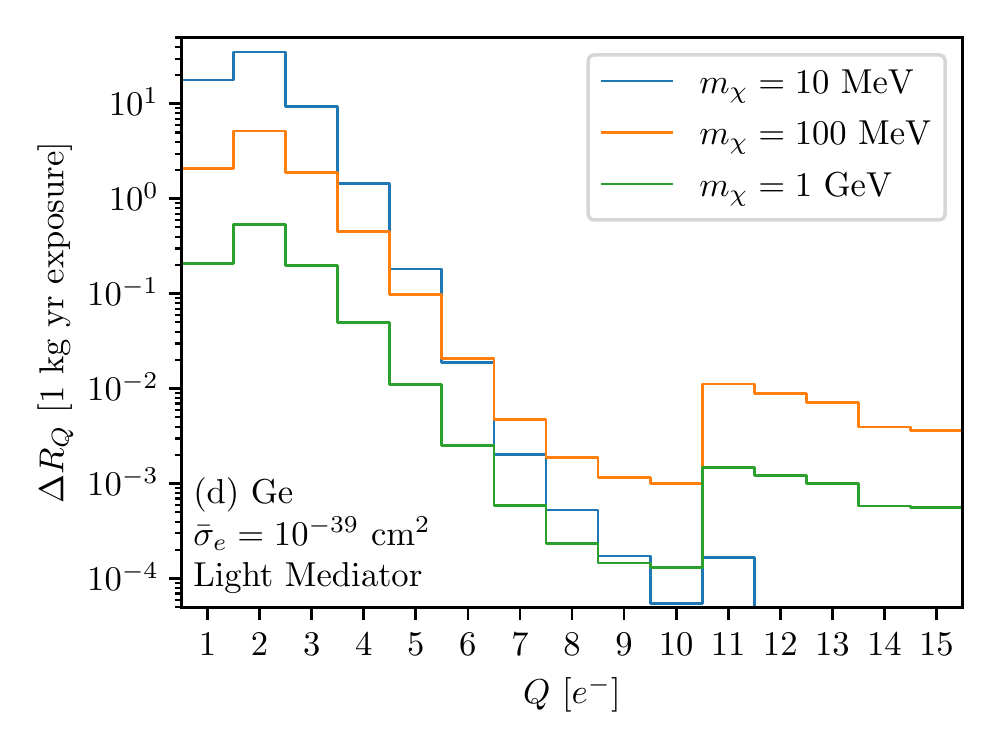}
            \caption{Panels (a) and (b) show the DM--electron scattering rates in a silicon crystal for heavy and light mediators respectively. These values are calculated using a TZP basis set with PBE0 exchange and correlation functionals, $4\times4\times4\ \mathbf{k}-$grid, and $q_\mathrm{max} = 25\alpha m_e$. Panels (c) and (d) show the DM--electron scattering rates in a germanium crystal with interaction mediated by a heavy and a light mediator respectively calculated using a def2-TZVP basis set with PBE0 exchange--correlation functional, $6\times6\times6\ \mathbf{k}-$grid, and $q_\mathrm{max} = 20\alpha m_e.$}
            \label{fig:mX}
        \end{figure*}
        \begin{figure*}
            \centering
            \includegraphics[width=0.497\linewidth]{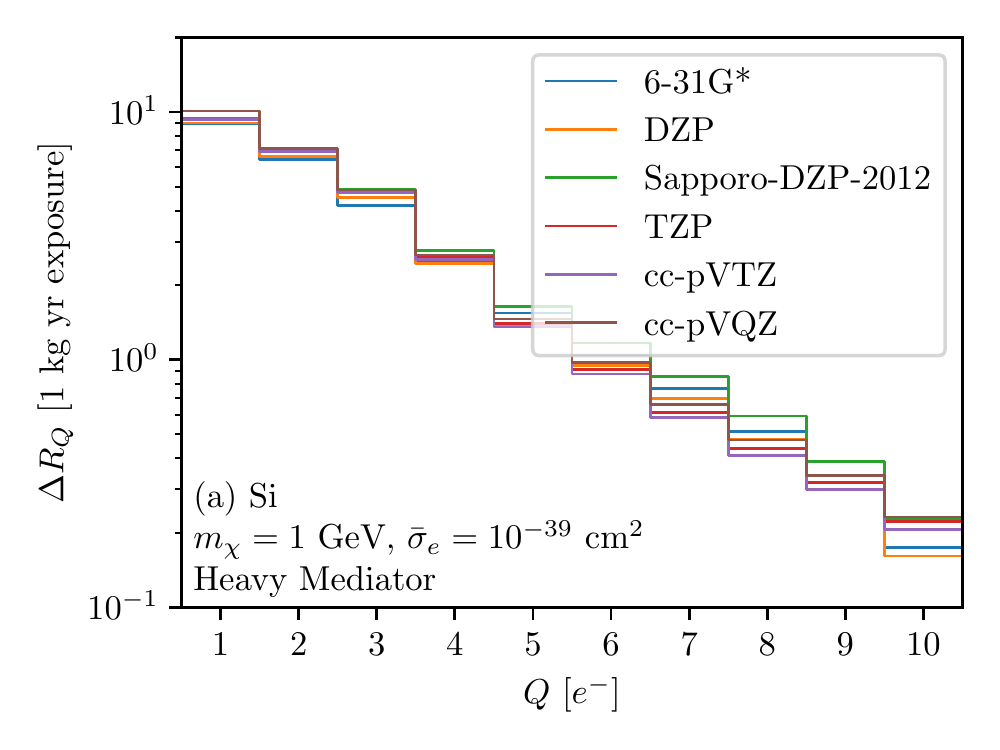}
            \includegraphics[width=0.497\linewidth]{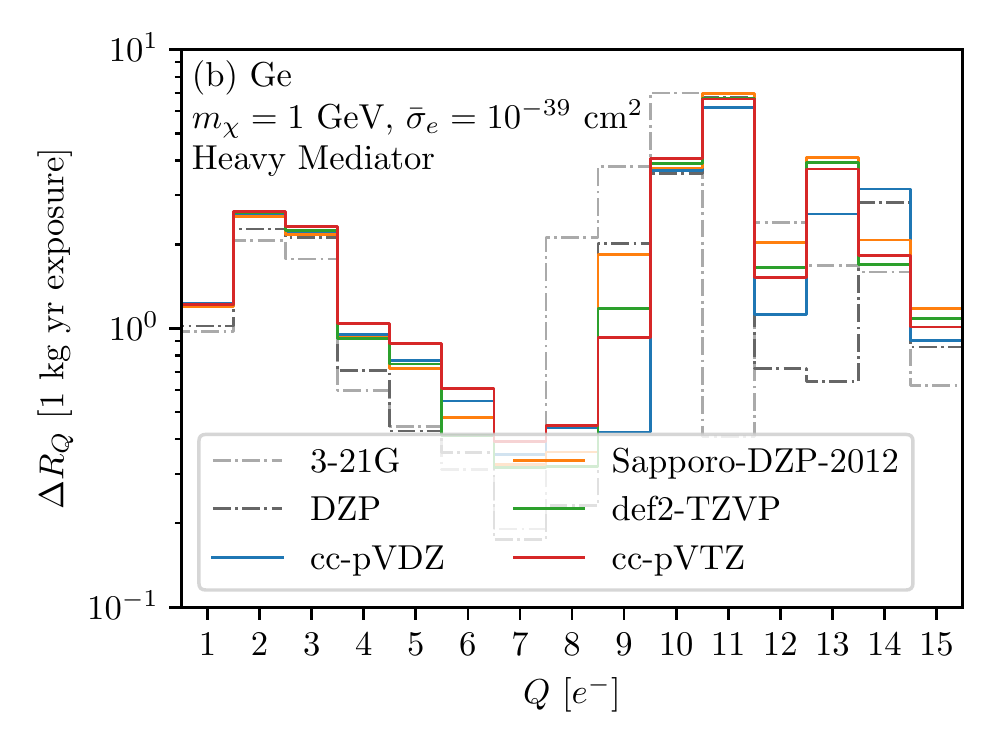}
            \caption{Panels (a) and (b) show the DM--electron scattering rates for Si and Ge respectively, calculated with a $4\times4\times4\ \mathbf{k}-$grid, with DM parameters noted in the figures. The plots use $q_\mathrm{max} = 10\ \alpha m_e$ and PBE exchange-correlation. \texttt{cc-pVQZ} is the most accurate basis set we test for Si, while \texttt{cc-pVTZ} is the same for Ge. Moving further, we use TZP and def2-TZVP for Si and Ge respectively, which mimic the most accurate bases at lower computational costs. }
            \label{fig:basis}
        \end{figure*}
        
        Fig.~\ref{fig:mX} shows the DM--electron scattering rates expected in silicon and germanium crystals for $m_\chi = 10$~MeV, 100~MeV, and 1~GeV for both heavy and light mediators, assuming the crystal form factors shown in Fig.~\ref{fig:cff}. In this and subsequent figures, we plot 
        \begin{equation}
            \Delta R_Q = \int dE_e\ \frac{dR}{dE_e} p(Q, E_e)\,,
        \end{equation}
        where $p(Q, E_e)$ is the probability that a transition with recoil energy $E_e$ excites $Q$ electrons (for more details, see \S\ref{subsec:second_ionization}). We use the ionization model at 100~K from~\cite{Ramanathan2020} for Si. For Ge, we use an electron-hole-pair creation model, 
        \begin{equation}\label{eq:pair_creation}
            Q = \sum_{n = 0}^\infty \Theta\left(E_e - n\times E_\mathrm{pp} - E_\mathrm{gap}\right),
        \end{equation}
        where $\Theta(x)$ is the Heaviside step function, $E_\mathrm{pp}$ is the electron-hole-pair creation energy ($E_\mathrm{pp} = 2.9$ eV for Ge), and $E_\mathrm{gap}$ is the bandgap of the material.

        Panels (a) and (b) of Fig.~\ref{fig:mX} show DM--electron scattering rates in a silicon crystal mediated by a heavy and a light mediator, respectively. For Si, we use a TZP basis set with a PBE0 exchange--correlation functional, $4\times4\times4\ \mathbf{k}-$grid and $q_\mathrm{max} = 25\alpha m_e$.
        
        Panels (c) and (d) of Fig.~\ref{fig:mX} show DM--electron scattering rates in a germanium crystal mediated by a heavy and a light mediator, respectively, calculated using a def2-TZVP basis set with a PBE0 exchange--correlation functional, $6\times6\times6\ \mathbf{k}-$grid and $q_\mathrm{max} = 20\alpha m_e.$
        
        Because the 3d-dominated bands in Ge are flat in $\mathbf{k}-$space (i.e., they are highly localized in real space), we need a denser $\mathbf{k}-$grid to reduce the numerical noise in the (unbinned) rate spectra $dR/dE_e$. An ionization model for Ge akin to the model in~\cite{Ramanathan2020} for Si (which includes a Fano factor) remains unavailable, which would smooth out the numerical noise while calculating $\Delta R_Q$ (see \S \ref{subsec:second_ionization}). 
        
        In principle, increasing the density of the \textbf{k}-grid would reduce the noise, at the expense of computation time. However the rates, barring $\sim 10\%$ systematics coming from the numerical noise at high $Q$, are robust (see \S\ref{subsubsec:k-grid_uncertainty} and Fig.~\ref{fig:k_grid} for more details), and indicate that for $m_\chi \gtrsim \order{100\rm\ MeV}$ the rates are higher for $Q\geq11\ e^-$ than for lower $Q$ (for lower masses, interactions with high $q$ transfers are kinematically suppressed). This would imply that germanium-based detectors with relatively high thresholds can still probe significant regions of DM parameter space, assuming $m_\chi \gtrsim 100$~MeV and a heavy mediator.

        The effects of all--electron modes are visible for heavy mediators, and not as much for light mediators. This is because of an effective $\abs{F_\chi}^2 \propto 1/q^4$ suppression in the DM--electron scattering cross-section in the case of light mediators.
        
    \subsection{Evaluation of Systematic Uncertainties}\label{subsec:4.3}
        
        There are multiple sources of theoretical uncertainties as well as several convergence parameters (i.e., parameters that can be improved with more computational time) in our calculation of DM--electron scattering rates. The choice of the exchange--correlation functional, $E_\mathrm{xc}[n]$, in Eq.~\eqref{eq:KS} is a source of theoretical uncertainty.  The real-space cut-off for constructing our Bloch atomic orbitals, the size of the $\mathbf{k}-$grid, and the choice of $q_\mathrm{max}$ are convergence parameters. The choice of the atomic centered Gaussian basis set is both a convergence parameter (since increasing the number of basis functions allows us to model the conduction states better) and a theoretical uncertainty (since different basis sets are optimized for different types of calculations, be it molecular or periodic boundary conditions). In this section, we go through each of these choices and determine their effects on our DM--electron scattering rate calculation.
        \begin{figure*}
            \centering
            \includegraphics[width=0.495\linewidth]{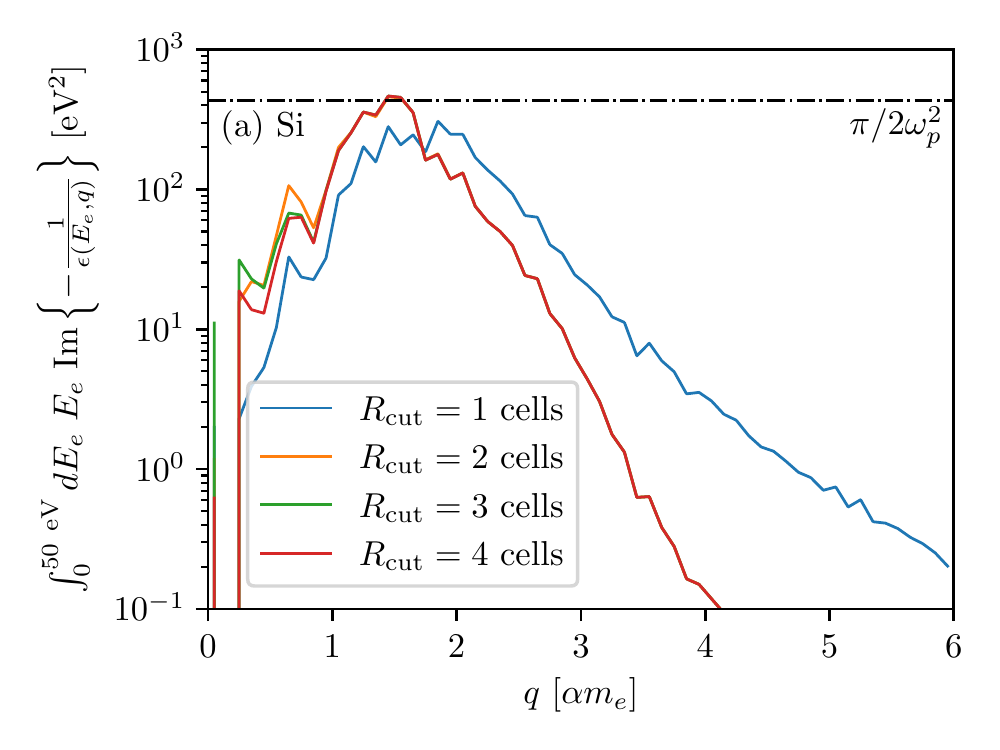}
            \includegraphics[width=0.495\linewidth]{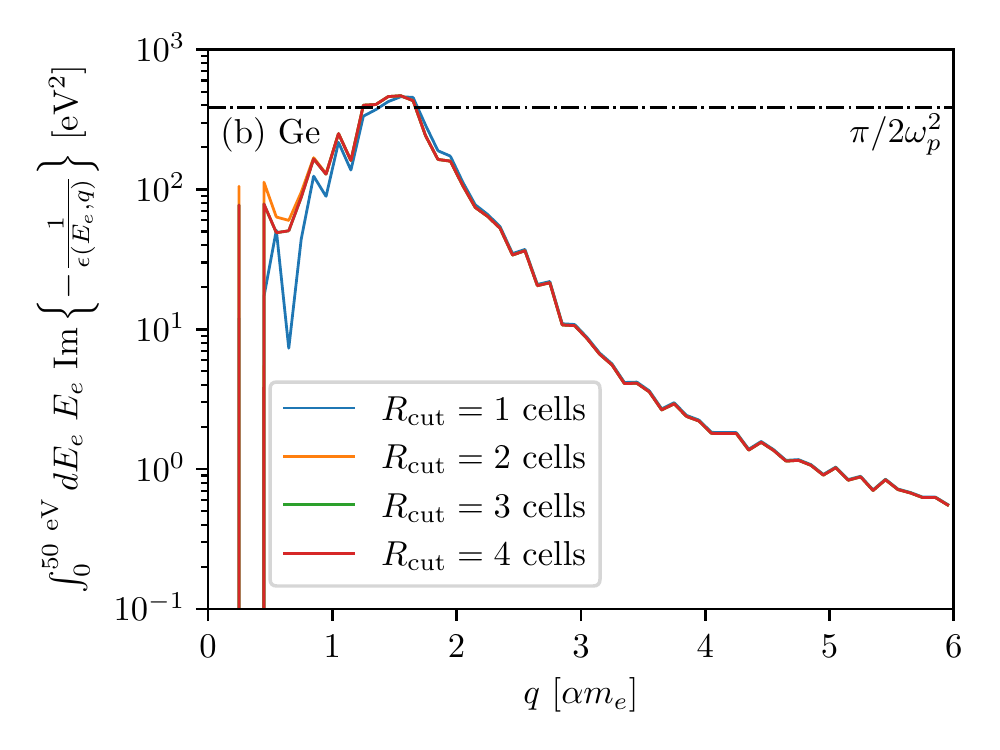}
            \caption{Panels (a) and (b) show the electron loss functions integrated over electron recoil energy, $E_e$, for Si and Ge, respectively, with the dash-dotted line showing the theoretical upper bound from the $f$-sum rule. For Si, we use here a TZP basis set, at $4\times4\times4$ \textbf{k}--grid and PBE functional to calculate these results. For Ge, we use here a def2-TZVP basis set, with the same \textbf{k}--grid and PBE functional. Note that a good description of high momentum transfer $q\gtrsim3\alpha m_e$ does not require a large $R_\mathrm{cut}$ for either element.\label{fig:sum_rule}}
        \end{figure*}
        \begin{figure*}
            \centering
            \includegraphics[width=0.497\linewidth]{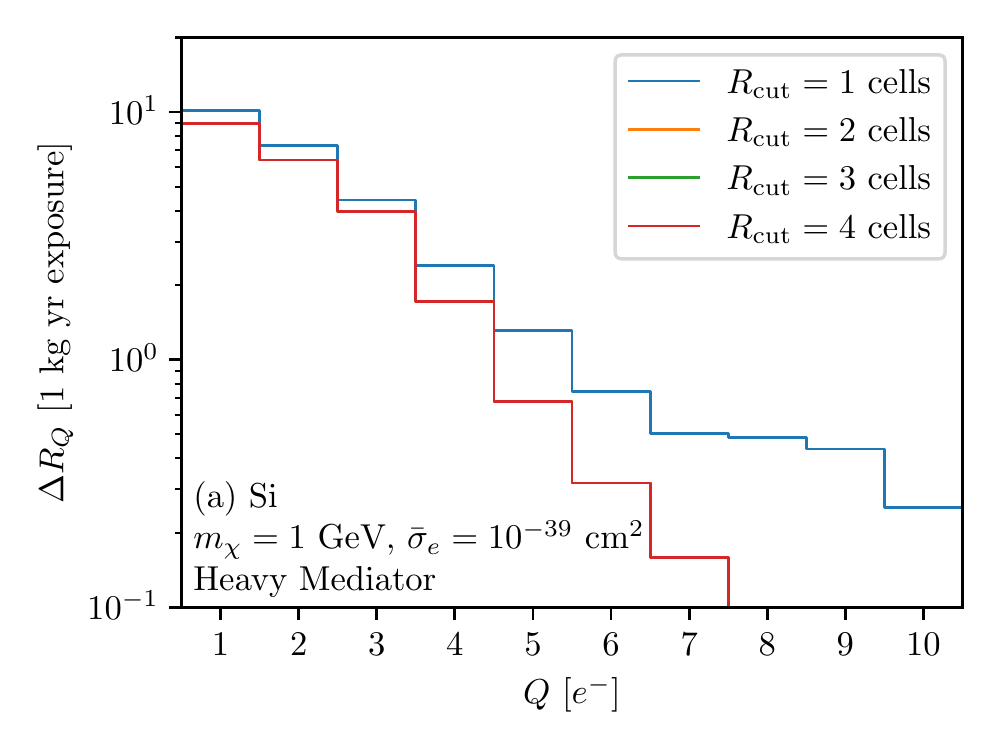}
            \includegraphics[width=0.497\linewidth]{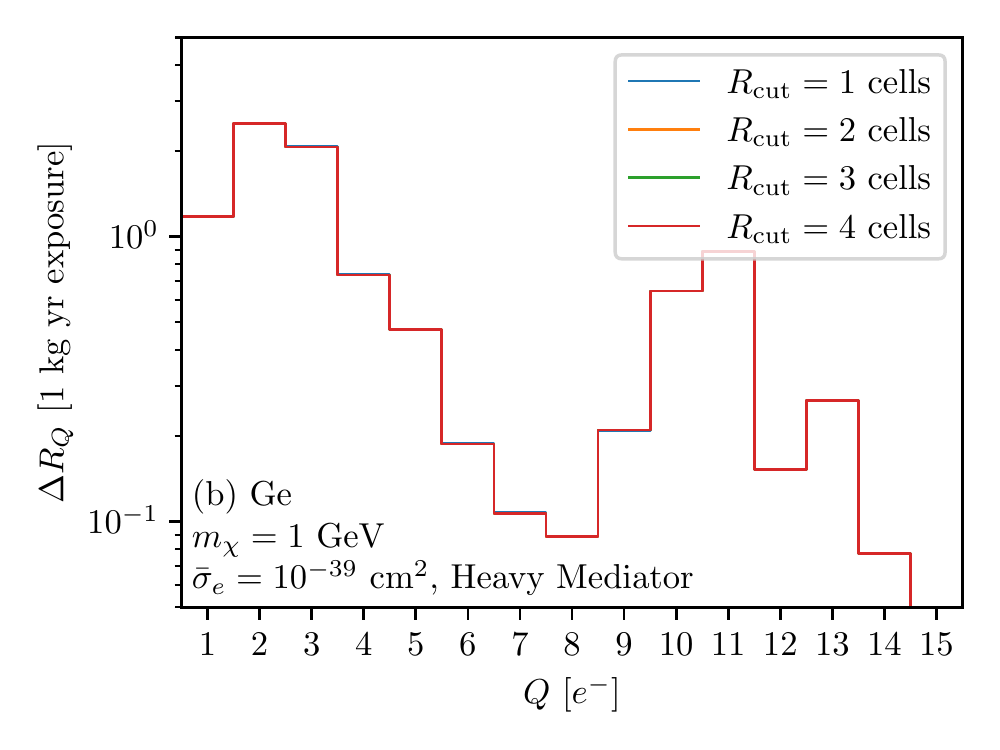}
            \caption{Panels (a) and (b) show the DM--electron scattering rates in Si and Ge, respectively, for various values of the real space cut-off $R_{\rm cut}$, assuming an exposure of 1~kg-year. We use a $4\times4\times4$ \textbf{k}--grid, set $q_\mathrm{max} = 6\alpha m_e$, and use the PBE functional.}
            \label{fig:Rcut_rates}
        \end{figure*}
        \begin{figure*}
            \centering
            \includegraphics[width=0.497\linewidth]{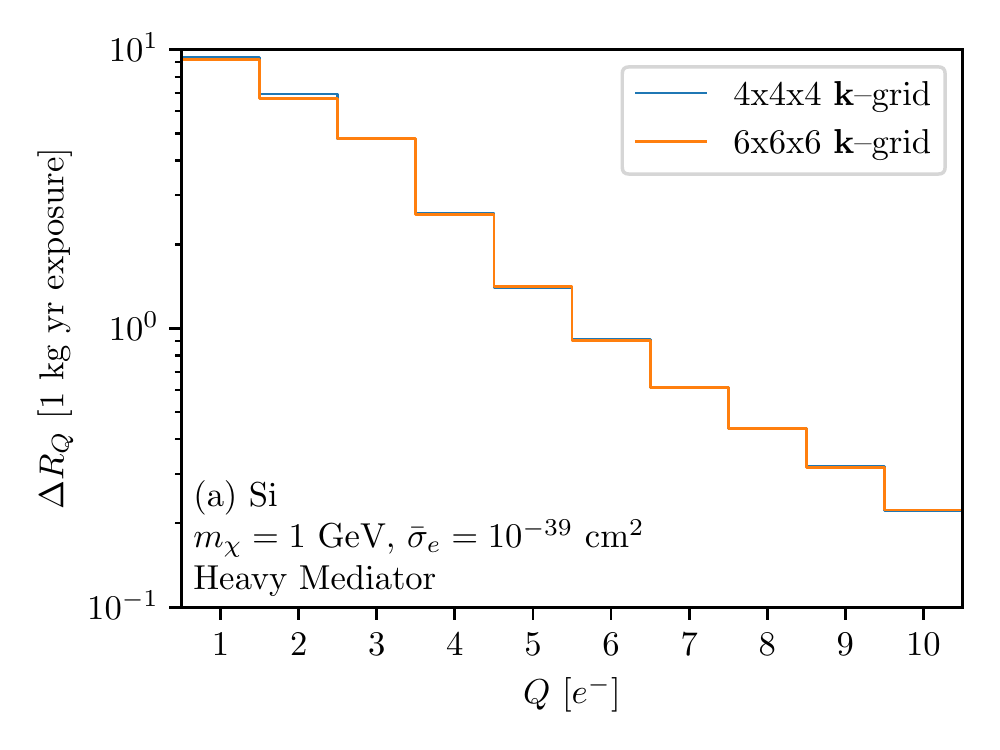}
            \includegraphics[width=0.497\linewidth]{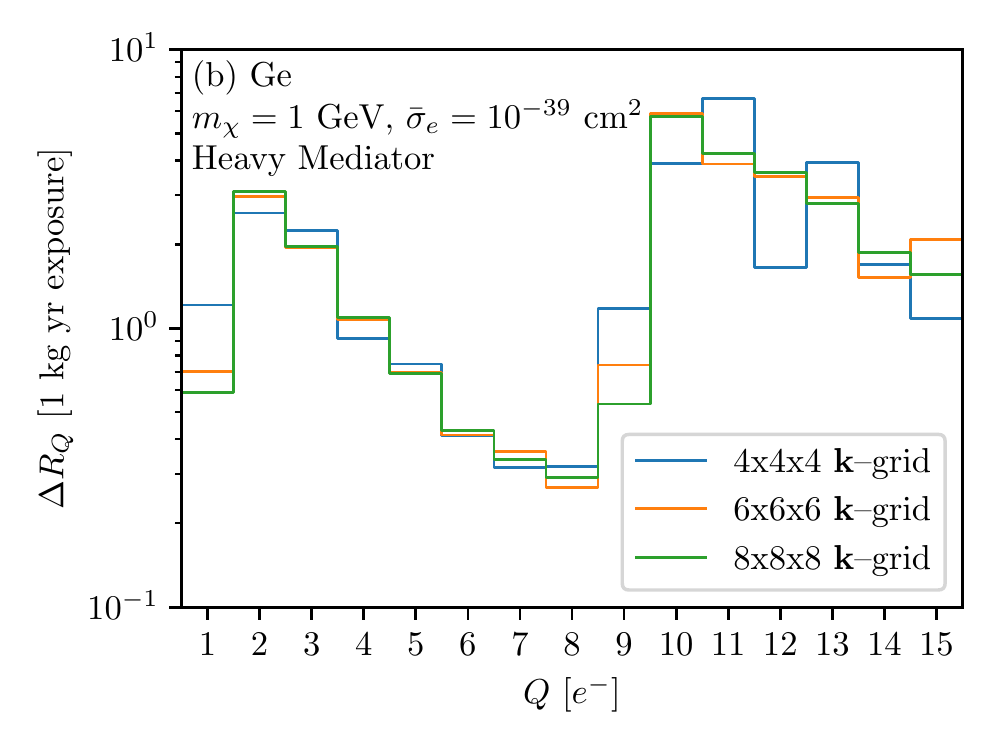}
            \caption{Panels (a) and (b) show DM--electron scattering rates for Si and Ge, respectively, for various $\mathbf{k}-$grid densities, assuming an exposure of 1~kg-year. We use the PBE exchange--correlation functional and set $q_\mathbf{max} = 10 \alpha m_e$. It is evident that the Si calculation is converged, even at the sparse $4\times4\times4\ \mathbf{k}-$grid level, while Ge only converges at higher $\mathbf{k}-$grid densities.}
            \label{fig:k_grid}
        \end{figure*}
        
        \begin{figure*}
            \centering
            \includegraphics[width=0.497\linewidth]{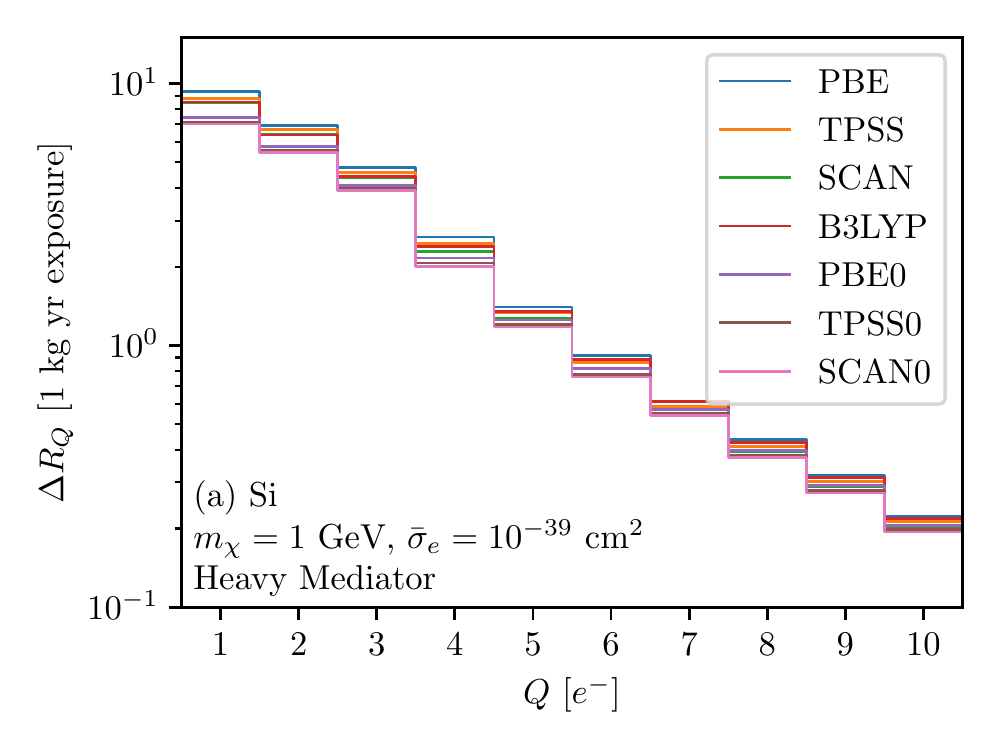}
            \includegraphics[width=0.497\linewidth]{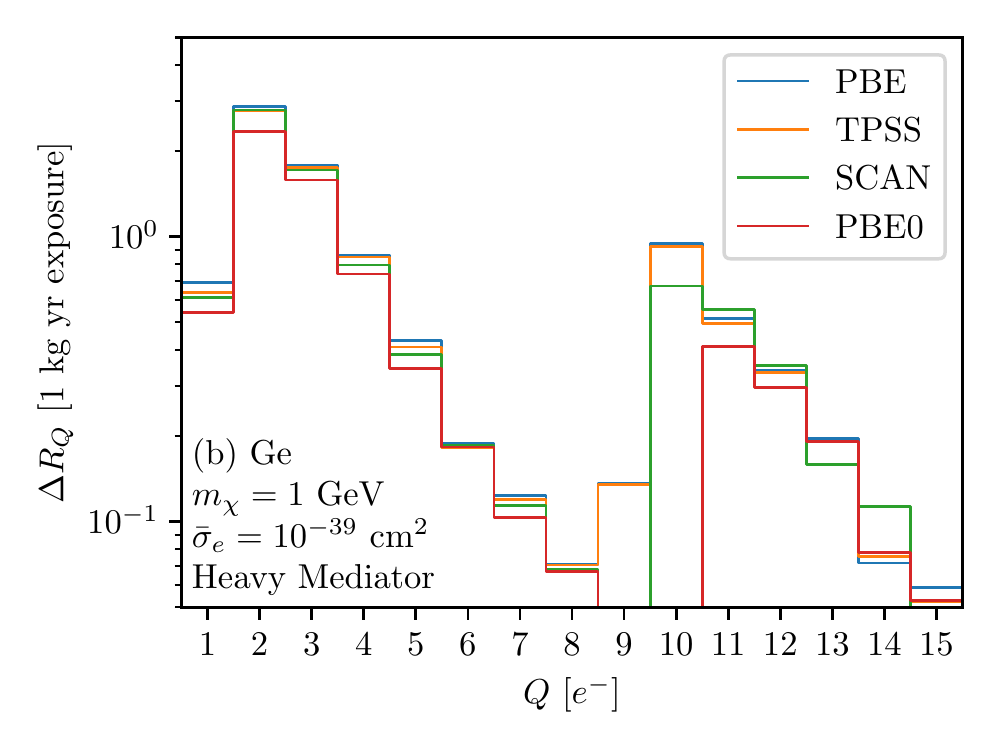}
            \caption{
            Panels (a) and (b) show DM--electron scattering rates for Si and Ge, respectively, calculated with various exchange-correlation functionals, assuming an exposure of 1~kg-year. We use a $4\times4\times4\ \mathbf{k}-$grid for Si and $6\times6\times6\ \mathbf{k}-$grid for Ge, and set $q_\mathrm{max} = 6 \alpha m_e$. We favor the well-tested PBE0 functional for our calculations.  Note the dependence of the energy of the 3d-shell of Ge on the choice of $E_{\rm xc}$.}
            \label{fig:xc}
        \end{figure*}
        \subsubsection{Choice of basis set}\label{subsubsec:basis_uncertainty}
        
        There are many choices of atom-centered Gaussian basis sets available for use \citep{BasisSetExchange}. However, most of these basis sets are optimized for molecular calculations, and we have to choose among the few optimized for periodic boundary conditions. In addition, the size of the basis sets determines the number of conduction bands.
        
        Fig.~\ref{fig:basis} shows the DM--electron scattering rates calculated for various basis sets. While the DM--electron scattering rates are consistent across all the basis sets we test, the cc-pVQZ (correlation-consistent polarized valence quadruple zeta) and cc-pVTZ (correlation-consistent polarized valence triple zeta) are the best optimized basis sets that we test for Si and Ge, respectively (these are also computationally very expensive). For Si, DM--electron scattering rates calculated using TZP (shown in Fig.~\ref{fig:mX}) differ by only $\sim5\%$ from those derived using cc-pVQZ. Similarly for Ge, DM--electron scattering rates calculated using def2-TZVP (shown in Fig.~\ref{fig:mX}) differ by only $\sim 5\%$ on average from those derived using cc-pVTZ. Both of the TZP and def2-TZVP basis sets provide a good balance of computational efficiency and accuracy, and we use these in further analyses. 
        
        Fig.~\ref{fig:basis} also shows that DM--electron scattering rates in Ge are heavily dependent on the choice of basis set, especially for large $Q \gtrsim 9\ e^-$. This is because basis sets like 3-21G and DZP are unable to capture conduction bands well, while the energy of the semi-core 3d electrons is highly dependent on accurate modelling of core shells.

        \begin{figure*}
            \centering
            \includegraphics[width=0.497\linewidth]{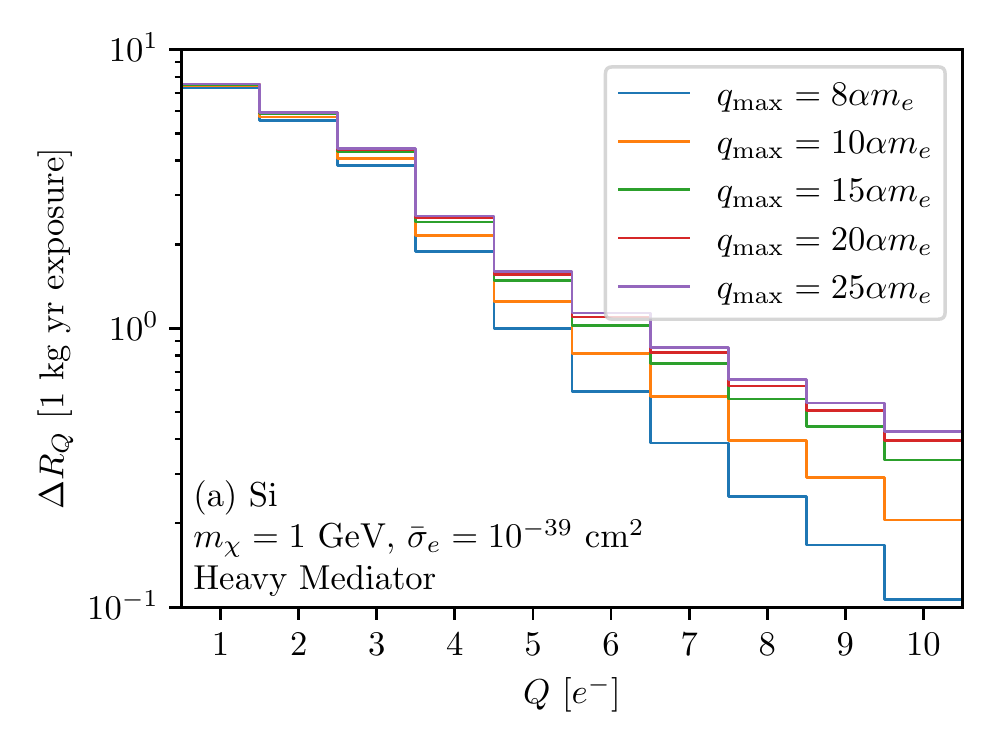}
            \includegraphics[width=0.497\linewidth]{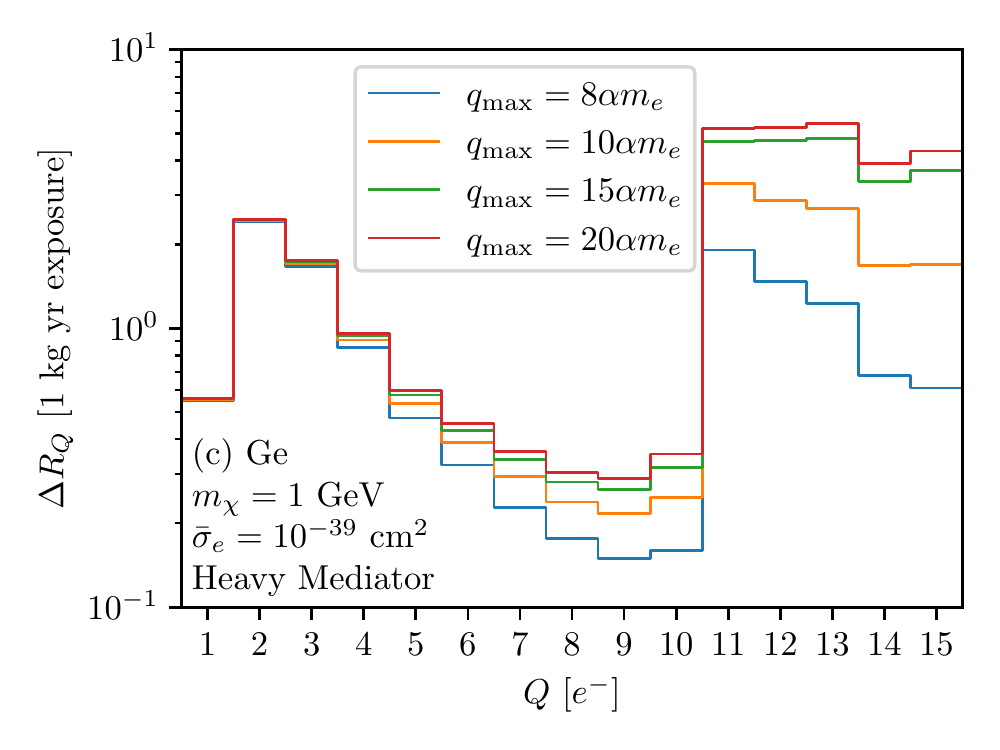}
            \includegraphics[width=0.497\linewidth]{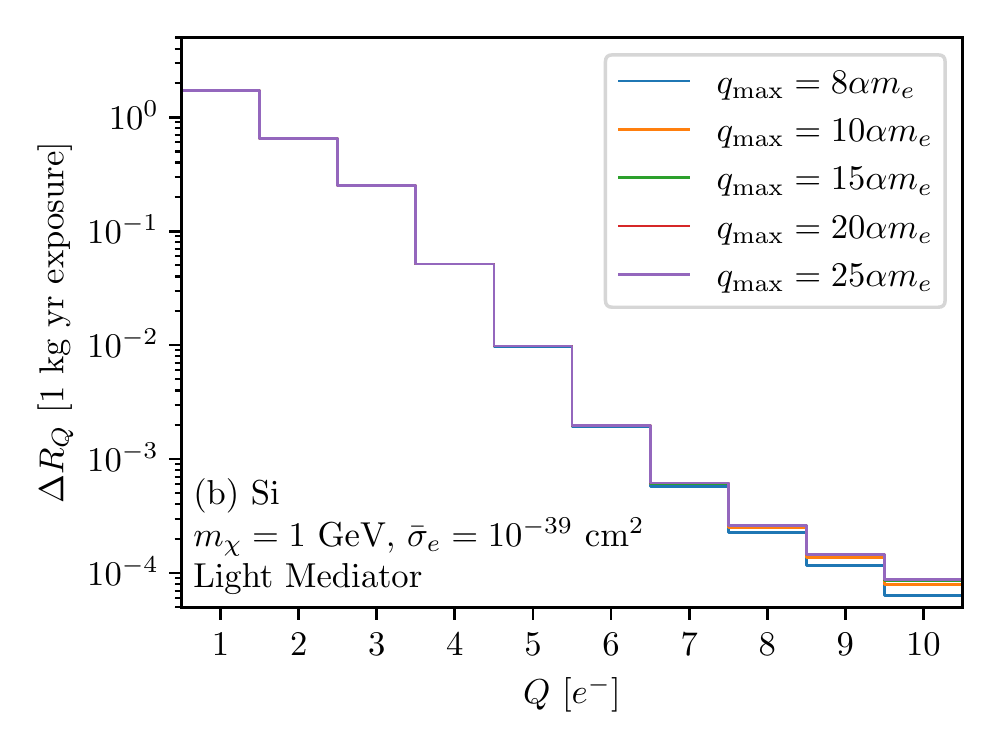}
            \includegraphics[width=0.497\linewidth]{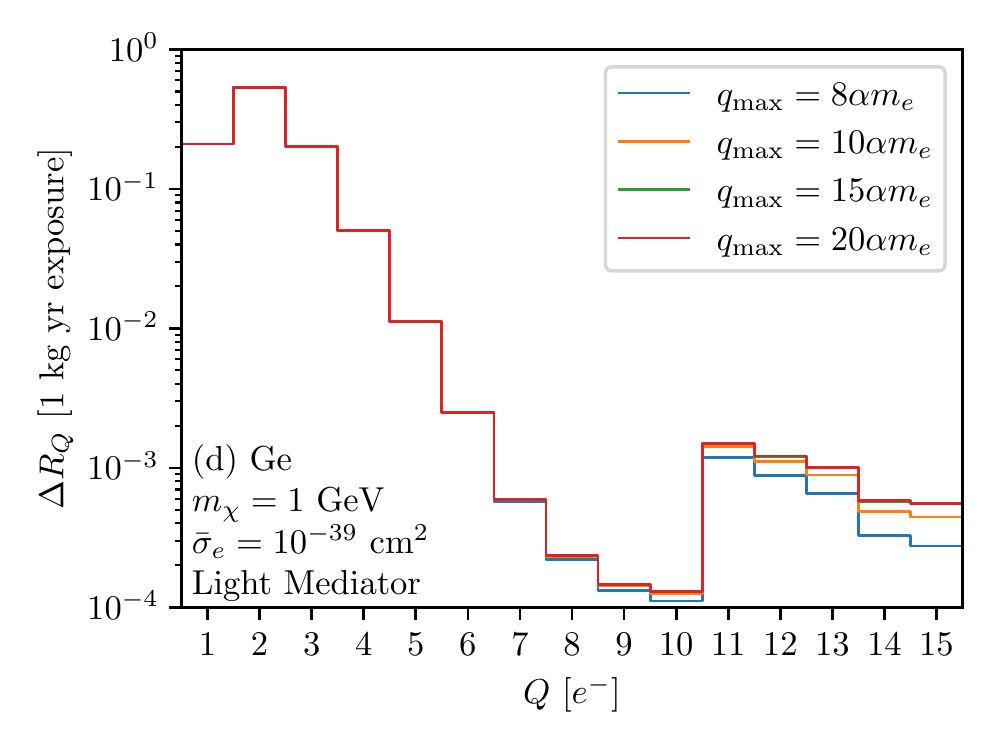}
            \caption{Panels (a) and (b) shows DM--electron scattering rates for Si with DM--electron interaction mediated by a heavy and a light mediator, respectively, with different $q_\mathrm{max}$ cutoffs, assuming an exposure of 1 kg-year. Panels (c) and (d) show the same for a Ge crystal. We use a $4\times4\times4\ \mathbf{k}-$grid for Si and a $6\times6\times6\ \mathbf{k}-$grid for Ge, and the PBE0 exchange correlation funtional.}
            \label{fig:qmax}
        \end{figure*}
        
        \subsubsection{Real space cutoff}\label{subsubsec:Rcut_uncertainty}

        Our atomic orbitals are Bloch sums in real space as in Eq.~\eqref{eq:blochsum}; in principle, one must sum over an infinite number of Gaussians displaced by real-space lattice vectors \textbf{R} to form each atomic orbital. In practice, however, Gaussians are rapidly decaying functions, and so it suffices to include a finite number of neighbors depending on the exponents in the contracted Gaussians. This is generally accomplished by setting a real space cut-off, which we call $R_\mathrm{cut}$.

        While PySCF is capable of choosing a dynamic real space cut off for each orbital, which lowers the computational cost to calculating the matrix elements (see Eq.~\ref{eq:mat_elems}), this complicates our analytical approach to calculating the matrix elements (see Appendix \ref{app:cart_gauss} for more details on an analytical calculation). Hence, we choose a constant $R_\mathrm{cut}$ for all orbitals, the value of which is chosen via the following procedure. We first calculate the electron loss function, i.e., the imaginary part of the inverse dielectric function $\text{Im}\left[-\epsilon(\omega,\textbf{q})^{-1}\right]$. For this, we assume that the real part of the dielectric function is modelled by Eq.~\eqref{eq:screening}, and calculate the imaginary part of the RPA dielectric function using Eq.~(16) of~\cite{Knapen:2021run}. We then integrate $\omega\text{Im}\left[-\epsilon(\omega,\textbf{q})^{-1}\right]$ over energy for a given magnitude of \textbf{q}; by the $f$-sum rule,
        \begin{equation}
            \int_0^\infty dE_e\ E_e \mathrm{Im}\left\{-\frac{1}{\epsilon(E_e, q)}\right\} = \frac{\pi}{2}\omega_\mathrm{p}^2
        \end{equation}
        the result should equal $\pi/2\omega_\text{p}^2$, where $\omega_\text{p}$ is the plasma frequency of the material. The $f$-sum rule is only achieved in limit of a complete basis set, however we have found that the convergence of this quantity is a useful diagnostic as to whether a given $R_\mathrm{cut}$ is sufficient for the relevant range of $q$~\citep{Hochberg2021Diel}. We show the electron loss function integrated over $E_e$ up to 50~eV for Si (using the TZP basis set) and Ge (using the def2-TZVP basis set) in Fig.~\ref{fig:sum_rule}.
        
        The high momentum transfer modes $q\gtrsim3\alpha m_e$ are captured well even by low $R_\mathrm{cut}$. One may understand this in the real space from the viewpoint of our KS wavefunctions -- the high frequency modes of the molecular orbitals, resulting from orthogonalization with inner orbitals, are more localised near the nuclei, and hence have smaller overlaps in real space with counterparts from more distant atoms. The low $q$ modes, on the other hand, correspond to long distance behavior of the matrix elements, and so necessitate the usage of larger $R_\mathrm{cut}$ in order to satisfy the $f$-sum rule. The curves going to zero at very low $q\lesssim0.3\ \alpha m_e$ is an artifact of our choice of a sparse $\mathbf{k}-$grid, and is not the true behavior of $f$ sum rule. 
        
        Motivated by these results, our final calculations use a hybrid real space cut-off. For Si (Ge), we use $R_\mathrm{cut} = 4\ (3)$ cells for $q\leq3\alpha m_e$ and $R_\mathrm{cut} = 2\ (1)$ cells otherwise. Fig.~\ref{fig:Rcut_rates} shows that this is also cautious, as the low momenta deviation only occurs in a prohibited region of the parameter space ($v_\mathrm{min}(q, E_e) > v_\mathrm{Escape} + v_\mathrm{Earth}$) for DM--electron scattering (see Appendix~\ref{app:derivation} for more details).

        \subsubsection{Convergence of $k-$mesh in reciprocal space}\label{subsubsec:k-grid_uncertainty}
        
        A potential source of systematic error in the rate calculation comes from the density of $k-$points in the first Brillouin Zone (1BZ). As discussed above, because the computational cost scales as the square of the number of $k-$points, it is infeasible to include $N_T$ $k-$points, where $N_T\gtrsim\order{10^{23}}$ is the number of unit cells in the crystal. In this section, we discuss the effects of modelling the 1BZ with an $N\times N\times N$ $\mathbf{k}-$grid, with a total of $N^3$ $k-$points.
        
        Fig.~\ref{fig:k_grid} shows the convergence of our calculations with $\mathbf{k}-$grid for both Si and Ge. Note that Si is already converged at a $4\times4\times4\ \mathbf{k}-$grid. One reason for this is that the numerical uncertainties are smoothed out from applying the ionization model (see \S\ref{subsec:second_ionization} for more details). For Ge, accurately describing transitions from the relatively dispersionless 3$d$-derived band at $E_e \sim 29$ eV below the Fermi level to the conduction bands requires a finer $\mathbf{k}-$grid. We find that a $6\times6\times6$ grid performs adequately, with errors of  $\lesssim 15\%$ compared to $8\times8\times8$ in each bin. A probabilistic ionization modelling for Ge, akin to~\citenum{Ramanathan2020} for Si, will aid in smoothing out the recoil spectrum.

        \subsubsection{Exchange-correlation functional}\label{subsubsec:XC_uncertainty}
        
        Fig.~\ref{fig:xc} shows DM--electron scattering rates in Si and Ge crystals for various exchange-correlation functionals. We test the commonly used PBE GGA functional, along with SCAN and TPSS mGGAs. For Si, we test multiple hybrids -- PBE0, SCAN0 and TPSS0, along with a hybrid semiempirical functional optimized for molecules rather than crystals (B3LYP). For Ge, we test PBE, SCAN, TPSS, and PBE0. 
    
        It is important to note that a scissor correction has been applied to the band gaps of Si and Ge, so DM--electron scattering calculations have the same gap regardless of functional. For materials where the experimental gap is not known, the differences in gaps predicted by different functionals is expected to lead to a significant source of variation in the scattering rates. A related issue observed for Ge (right panel of Fig.~\ref{fig:xc}) is the dependence of the energy of the 3d shell. This results in significant differences in the DM--electron scattering rates in the 8-11 electron-hole-pair bins. It is apparent that PBE, TPSS, and SCAN underestimate the electron binding energy for the 3$d$-shell electrons, with values $\sim$25~eV from the top of the valence band. The PBE0 functional results in values between 28.6 and 29.0~eV, which are much closer to the experimental values of $\sim$29.5~eV of 3d-shell electrons, respectively~\cite{Bearden1967}. For Si, we scissor correct the bandgap, and core orbitals do not get involved until energies of $\sim 99.2$ eV~\cite{Bearden1967}.

        \subsubsection{Maximum momentum transfer}\label{subsubsec:qmax_uncertainty}
        
        The implementation of atom-centered basis sets without using an effective core potential has one direct effect---we are able to capture the high-momentum transfer regime of the crystal form factor. These high-$q$ contributions come from orthogonalizing the valence and conduction bands against the core orbitals, which introduces high wavenumber modes to the valence and conduction wavefunctions. This allows the wavefunctions to be modelled to arbitrarily high wavenumbers, and allows us to fully capture the crystal form factor. Fig.~\ref{fig:qmax} shows the impact of adding high $q$ modes on the rates of 1 GeV DM particle interacting with Si (left panels) or Ge (right panels)  via a heavy (top row) or light (bottom row) mediator. 
        
        As expected, DM--electron scattering mediated by a light mediator is not significantly influenced by the high-$q$ contributions in silicon. This is due to the $\abs{F_\chi (q)}^2 \propto q^{-4}$ dependence of the rate in the integrand of Eq.~\eqref{eq:finRate}. Ge, on the other hand, is sensitive to $q_{\text{max}}$ even for scattering through a light mediator, since the 3$d$-shell in germanium dominates the high $q$ regime. 
        
        For interactions mediated by a heavy boson, however, there are important high-$q$ contributions even for relatively small charge bins. Moreover, when including the high-$q$ contributions, we see that DM with $m_\chi \gtrsim 50$ MeV and scattering through a heavy mediator ($F_\chi \approx 1$), the rates from $Q \geq 11\ e^-$ bins dominate over the $1\ e^-\leq Q\leq 10\ e^-$ bins. Similarly, for Si with a heavy mediator, high $q$ contributions are important, with, e.g., a $\sim 75\%$ increase in rates if we go from $q_\mathrm{max} = 8\alpha m_e$ to $q_\mathrm{max} = 25\alpha m_e$ for the $10\ e^-$ bin.
    
    \subsection{Effects of the secondary ionization model for silicon}\label{subsec:second_ionization}
    
\begin{figure}[t]
        \centering
        \includegraphics[width=\linewidth]{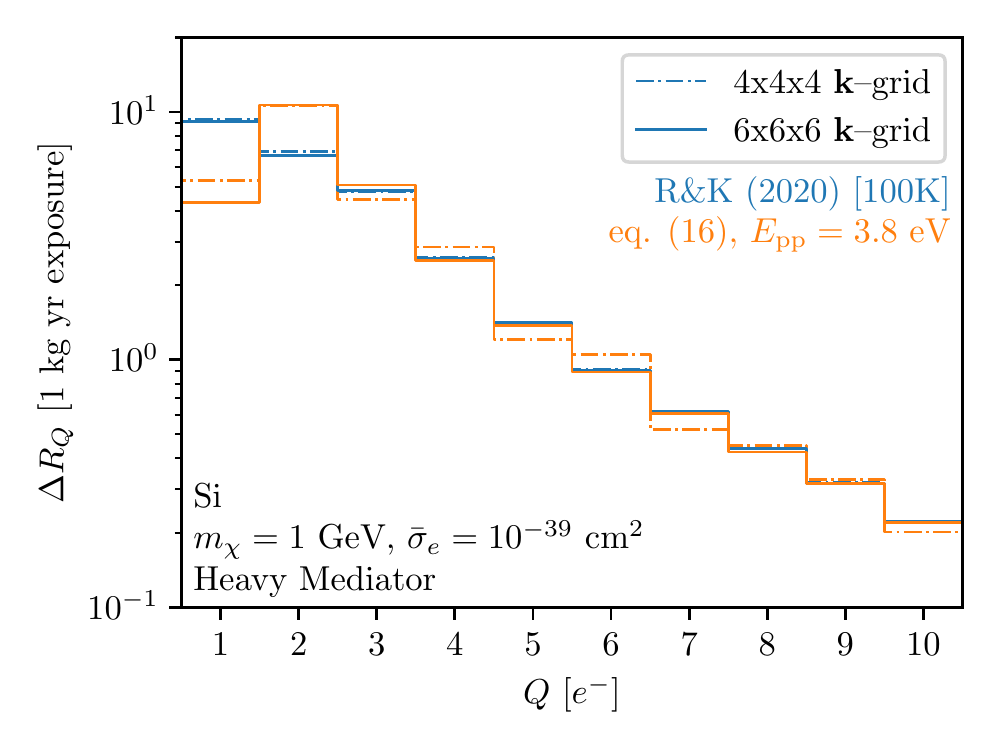}
        \caption{The effects on the DM-electron scattering rates in silicon of using the secondary ionization modeling from~\cite{Ramanathan2020} (``R\&K'') versus the step-function model from Eq.~\eqref{eq:pair_creation}. We use the PBE exchange-correlation functional and $q_{\rm max}= 10\ \alpha m_e$.}
        \label{fig:second_ionization}
    \end{figure}

    Our results for the DM-electron scattering rates in silicon are shown using the ionization modelling from~\cite{Ramanathan2020}.  In Fig.~\ref{fig:second_ionization}, we compare these rates with those from a simple step function model from Eq.~\eqref{eq:pair_creation} for Si. For the latter, we use $E_\mathrm{gap} = 1.1$ eV and $E_\mathrm{pp} = 3.8$~eV (see, e.g., \cite{Klein:1968}). 
    We see significant differences between the two ionization models for the 1 $e^-$ and 2 $e^-$-bin, although the rates are similar for the bins with $Q\geq3$. As we observed for Ge, for which only a step-function model is available, the probabilistic model from~\cite{Ramanathan2020} smoothes out the numerical fluctuations introduced by the sparse $\mathbf{k}-$grid.
        \begin{figure*}
        \centering
        \includegraphics[width = \linewidth]{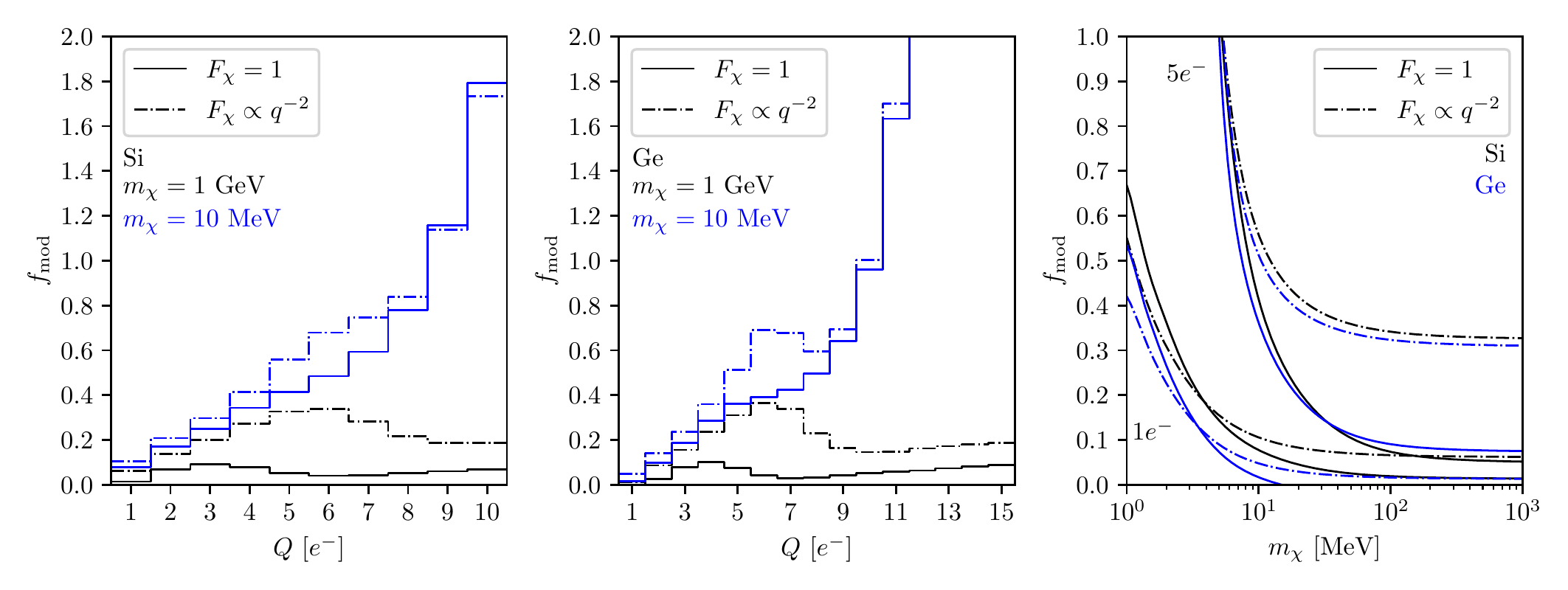}
        \caption{The modulation amplitude, $f_\mathrm{mod}$ from Eq.~\eqref{eq:annual_mod}, versus $Q$ for Si (left) and Ge (middle), for $m_\chi=10$~MeV and 1~GeV, for both heavy and light mediator-mediated scattering.  The right panel shows $f_\mathrm{mod}$ versus $m_\chi$, for $Q = 1 e^- \text{ and }Q = 5e^-$, for both Si and Ge. 
        We use the PBE0 exchange-correlation functional, along with $q_{\rm max}= 25\ \alpha m_e\ (20\ \alpha m_e)$ and a $4\times4\times4\ (6\times6\times6)\ \mathbf{k}-$grid for Si (Ge).} 
        \label{fig:annual_mod}
    \end{figure*}
    \begin{figure*}
        \centering
        \includegraphics[width = 0.497\linewidth]{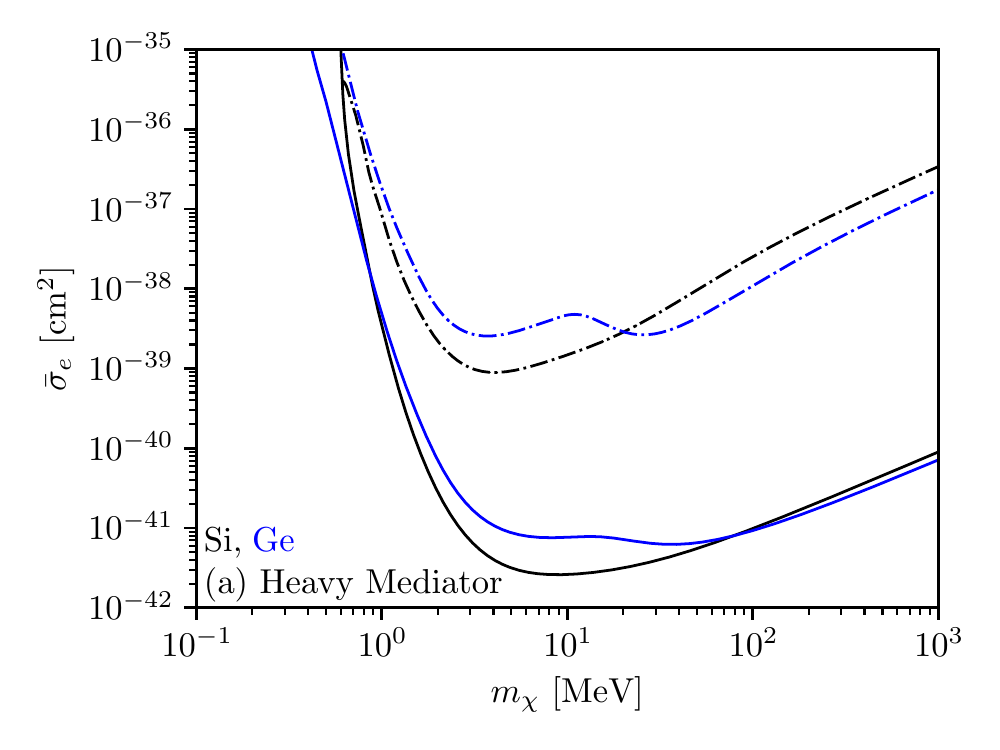}
        \includegraphics[width = 0.497\linewidth]{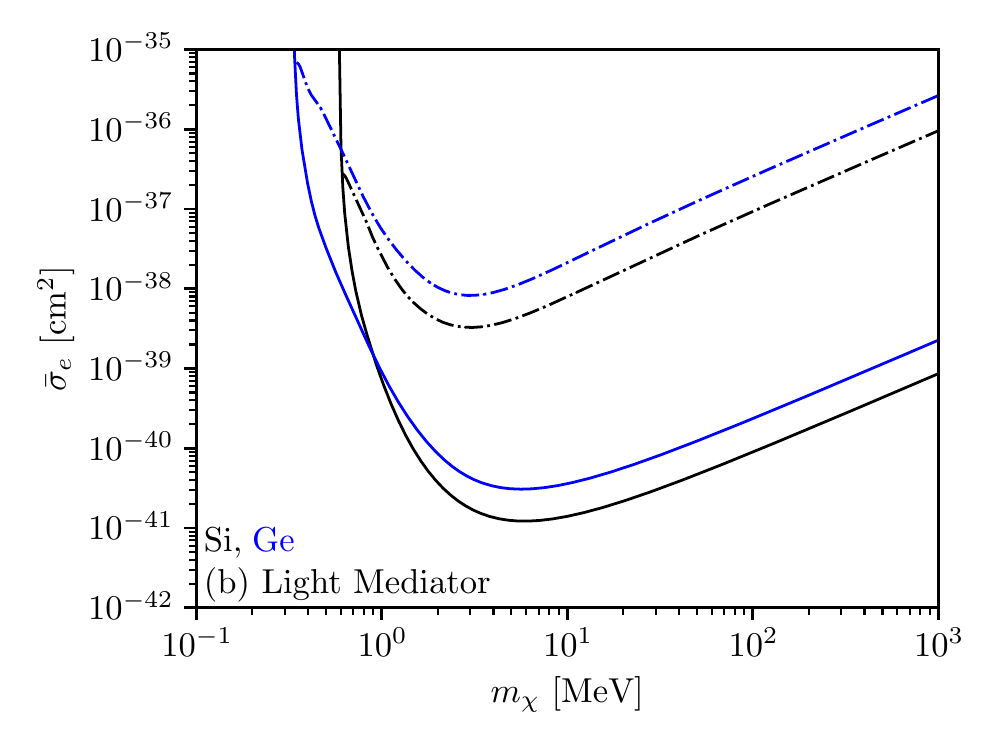}
        \caption{This plot shows the reach for both, obtaining 2.3 events for 1 kg$\cdot$yr exposure of our target material (solid line), as well as the threshold for a 5$-\sigma$ discovery by annual modulation with the same exposure (dash-dotted line). Panel (a) shows the upper bounds on cross-section that can be placed for heavy mediators, while panel (b) shows the same for light mediators. We assume no background for either panel, and include $1\leq Q\leq10$ $e^-$ for Si (black) and $1\leq Q\leq 15\ e^-$ for Ge (blue).}
        \label{fig:reach}
    \end{figure*}
    \subsection{Annual Modulation}

    \begin{figure*}
        \centering
        \includegraphics[width=0.497\linewidth]{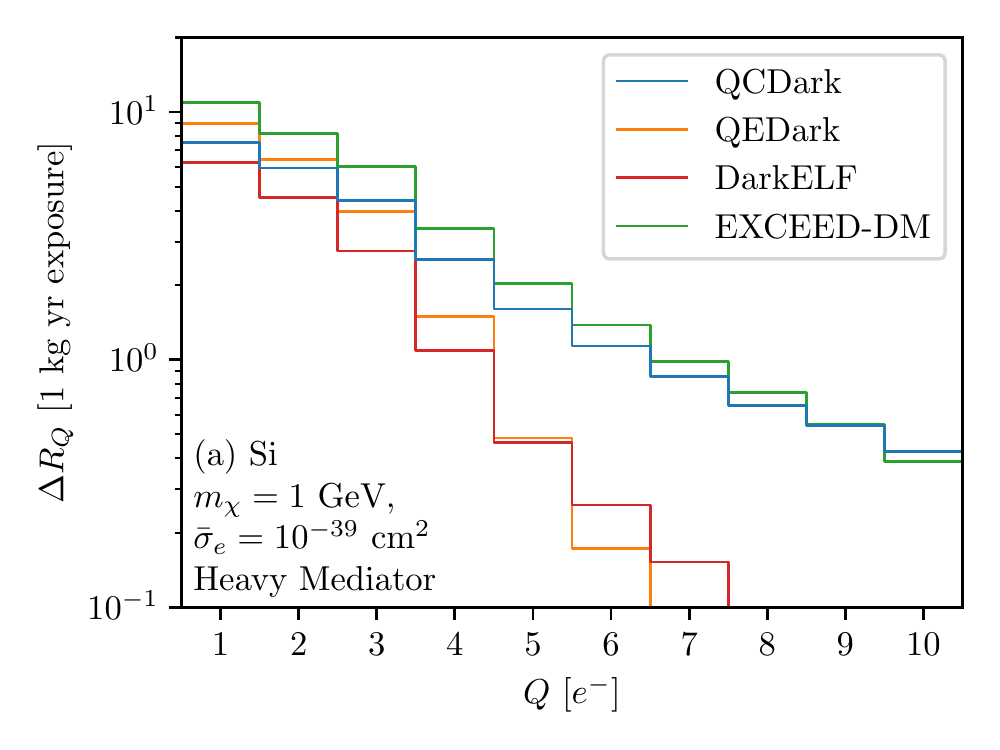}
        \includegraphics[width=0.497\linewidth]{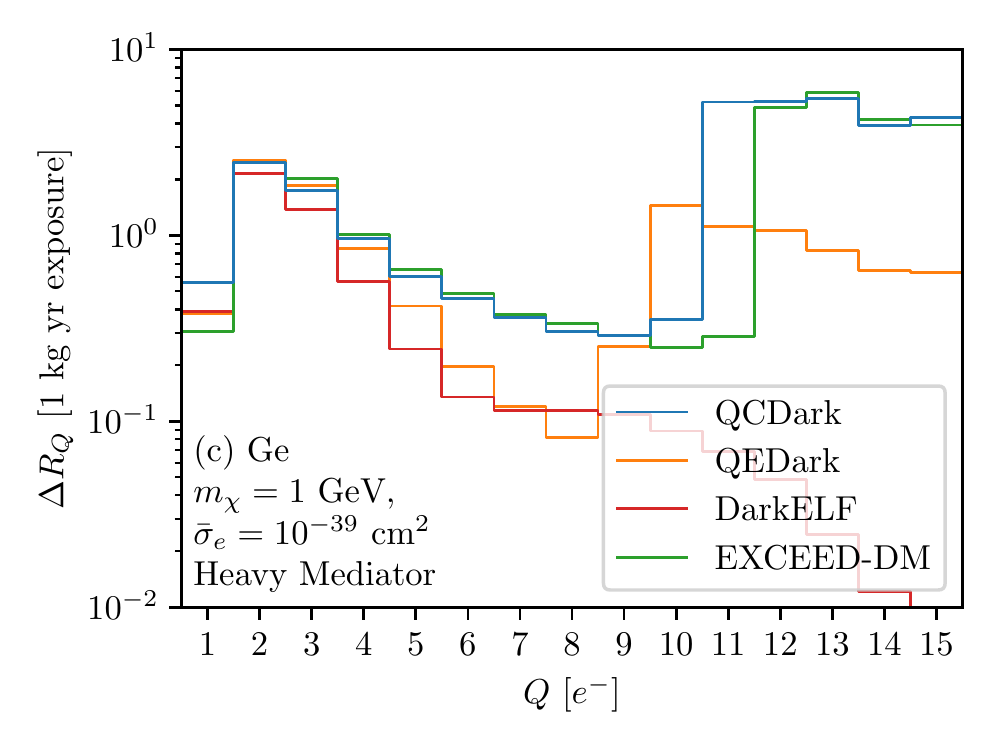}
        \includegraphics[width=0.497\linewidth]{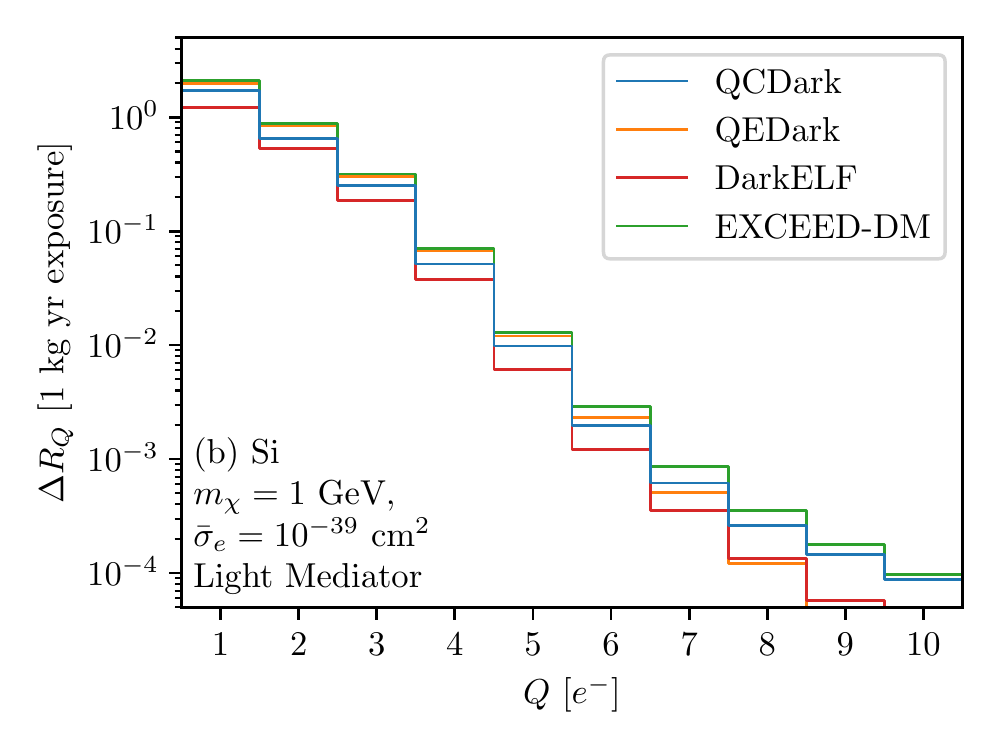}
        \includegraphics[width=0.497\linewidth]{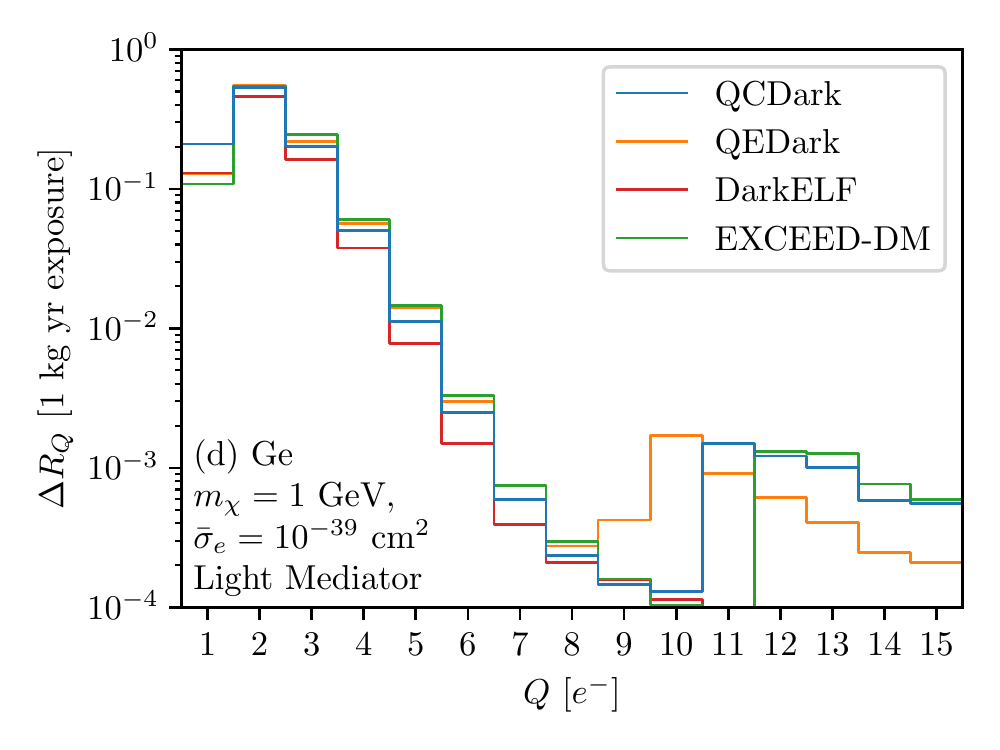}
        \caption{Panels (a) and (b) show the comparison among DM--electron scattering rates in Si assuming a heavy and a light mediator, respectively, calculated using different codes. This work and QEDark both implement the screening described in \S \ref{subsec:2.2.1}, while EXCEED-DM and DarkELF use their numerically calculated dielectric function. Panels (c) and (d) show the comparative plots for Ge assuming a heavy and a light mediator respectively.}
        \label{fig:comparison}
    \end{figure*}
    
    The DM--electron scattering rates are dependent on the DM flux incident on the target material, which in turn depends on the velocity of the detector in the galactocentric frame. For table-top experiments, there are three major contributions to this velocity.  First, there is the local circular velocity, which we take to be $v_0 = 230$ km s$^{-1}$. The second contribution comes from the Sun's peculiar velocity, $\mathbf{v}_\odot - \mathbf{v}_0$. We use the recommended value, $v_\odot = 250.2$ km s$^{-1}$~\cite{Baxter2021}. Finally, the earth revolves around the sun with an average speed $\expval{v_\oplus} = 29.8$ km s$^{-1}$. This revolution causes an annual modulation in DM--electron scattering rates, as the total velocity, $v_\mathrm{Earth} = \abs{\mathbf{v}_\odot + \mathbf{v}_\oplus}$, varies from $220.4$ km s$^{-1}$ on December 2 to $280$ km s$^{-1}$ on June 2. 
    
    We calculate the modulation amplitude following~\cite{Essig:2015cda},
    \begin{equation}\label{eq:annual_mod}
        f_\mathrm{mod}^Q = \frac{\Delta R_{Q, \mathrm{Jun 2}} - \Delta R_{Q, \mathrm{Dec. 2}}}{2\Delta R_{Q, 0}}\,,
    \end{equation}
    where $\Delta R_{Q, 0} = \Delta R_{Q, \mathrm{Sept. 2}} = \Delta R_{Q, \mathrm{Mar. 2}}$. Even in the presence of backgrounds, a measurement of $f_\mathrm{mod}$ could allow for the detection of DM in such an experiment. 
    We plot $f_\mathrm{mod}$ as a function of the DM mass, the mediator form factor, and the charge ionized in the target material $Q$ for both Si and Ge in Fig.~\ref{fig:annual_mod}. 
    
    Comparing to Fig.~8 of~\cite{Essig:2015cda}, the difference in rates from the inclusion of high wavenumber modes in the crystal form factor allows the electron to scatter into a larger parameter space, which reduces $f_{\rm mod}$, especially for the $m_\chi = 1$ GeV case. The same is visible for Ge, and is in fact even more pronounced for the 3d-shells, which dominate the rate.
    
    A measurement of the annual modulation signal will be an important step in confirming a potential DM signal.  We calculate the 5$\sigma$-sensitivity by requiring 
    \begin{equation}
        \frac{\Delta S}{\sqrt{S_\mathrm{tot} + B}}\geq 5,
    \end{equation}
    where $\Delta S = f_\mathrm{mod}S_\mathrm{tot}$ is the modulation amplitude, $S_\mathrm{tot}$ is the total number of signal events, and $B$ is the number of background events. Here $f_\mathrm{mod}$ is calculated using Eq.~\eqref{eq:annual_mod}, except we sum the rates over $1\ e^-\leq Q\leq 10\ e^-$ for Si, and $1\ e^-\leq Q\leq 15\ e^-$ for Ge. Assuming no background events, we show the $5\sigma$ discovery reach in Fig.~\ref{fig:reach} with dash-dotted lines for Si (black) and Ge (blue). The left panel shows the reach for heavy mediators, while the right panel corresponds to light mediators.
    
    \subsection{Comparison with other codes}\label{sec:5}
    
    \texttt{Quantum Espresso} was used in the first numerical calculation of the crystal form factor for Ge in~\cite{Essig:2011nj}.  It was also used in~\cite{Essig:2015cda}, which presented a detailed calculation of the crystal form factor for both Ge and Si, and made the resulting code, \texttt{QEDark}, publicly available.  In Fig.~\ref{fig:comparison}, the crystal form factor is recalculated with \texttt{QEDark} with improved computational parameters, including a higher energy cutoff for plane wave calculations, a denser $\mathbf{k}-$grid for both Si and Ge, and the analytical screening described in \S \ref{subsec:2.2.1}. We use the PBE functional, which, as shown above, underestimates the energy of the 3d-shell in Ge, and also excludes the effects of high frequency modes that are visible in both Si and Ge, especially for heavy mediators. 
    
    DarkELF~\cite{Knapen2021Diel} emphasized the need for better screening, especially for low energy excitations. Here we use GPAW RPA dielectric function with Local Field Effects (LFE) for both Si and Ge. However, it also does not include the high frequency modes, which dominate the rates at high energy, and the Ge 3d-shell is frozen in the pseudopotential, which otherwise dominate the DM--electron scattering rates at $E_e\gtrsim 30$ eV. 
    
    EXCEED-DM~\cite{Trickle2022} is able to reconstruct the high-frequency modes, and is also able to capture the dielectric screening with an RPA dielectric function. It also employs the well-tested and commonly employed HSE06 functional. However, it reconstructs the semi-core and core orbitals after a pseudopotential calculation. We find good agreement between the rates calculated with EXCEED-DM and \texttt{QCDark}, showing that the PAW method is accurate in these materials.
    
    \texttt{QCDark} implements \textit{ab-initio} calculation of the crystal form factor, along with an analytical approximation to the dielectric function. However, as~\cite{Trickle2022} recently showed, the analytic screening only approximates the true screening, and does not capture all the effects completely. On the other hand, \texttt{QCDark} allows for a much better handle on systematics by giving users control over the theory parameters, as discussed in \S\ref{subsec:4.3}. 

\section{Conclusion}\label{sec:6}
    In this paper, we present DM--electron scattering rates in silicon and germanium crystals calculated using a new code, which we make public as \texttt{QCDark}. We use a novel approach that naturally includes all core electrons, and treats them on the same level as valence electrons of the crystal. This implies that all-electron effects are automatically included from the beginning. Moreover, we present a systematic treatment of the theoretical uncertainties associated with the calculation, including those associated with DFT (basis set, exchange--correlation functional, and $\mathbf{k}-$grid), along with uncertainties associated with the transition matrix elements (real space cutoff and the maximum momentum transfer modelled, $q_{\rm max}$). 
    
    The major sources of systematic error include the choice of basis set and exchange--correlation functional, even after we apply the scissor correction, and the choice of $q_\mathrm{max}$, though the rates converge quickly in the $E_e\in[0, 50\text{ eV}]$ range for both Si and Ge. The rates also converge quickly as finer $\mathbf{k}-$grids are chosen, assuming the secondary ionization model in~\cite{Ramanathan2020} for Si. The rates also converge quickly for small values of the real-space cutoff, $R_\mathrm{cut}$.
    
    We find that modelling high momentum transfers by including all-electron effects is necessary for accurately modelling DM--electron scattering rates, especially at high recoil energies, in line with the findings in~\cite{Griffin21, Trickle2022}. This is especially important for Ge crystals, in which the transition rates from the 3d-shell (when kinematically accessible) dominate the rates if the high momentum transfer modes are modelled accurately.

\section*{Acknowledgements}
    We thank Daniel Baxter, Timothy Berkelbach, Yonit Hochberg, Simon Knapen, Yutaro Shoji, Greg Suczewski, Tanner Trickle, and Tien-Tien Yu for valuable discussions. C.E.D.~acknowledges support from the National Science Foundation under Grant No.~DMR-2237674. The Flatiron Institute is a division of the Simons Foundation.  R.E.~acknowledges support from DoE Grant DE-SC0009854, Simons Investigator in Physics Award 623940, and the US-Israel Binational Science Foundation Grant No.~2016153. A.S.~and C.Z.~were supported in part by a Stony Brook IACS Seed Grant, from Fermilab subcontract 664693 for the DoE DMNI award for Oscura, from DoE Grant DE-SC0009854, and from the Simons Investigator in Physics Award 623940. We also thank Stony Brook Research Computing and Cyberinfrastructure, and the Institute for Advanced Computational Science at Stony Brook University for access to the high-performance SeaWulf computing system, which was made possible by a National Science Foundation grant No.~1531492.

\bibliography{bib}

\clearpage 
\appendix

\section{Properties of Cartesian Gaussians}\label{app:cart_gauss}

    In this section we discuss the properties of Cartesian Gaussians, including the calculation of the matrix elements in Eq. \eqref{eq:mat_elems}. Our Cartesian Gaussian basis sets contain primitive Gaussians as building blocks (see Eq.~\eqref{eq:cartestian_gauss}), 
    \begin{equation}
    \begin{split}
        G_{ijk} (\mathbf{r}, \xi, \mathbf{A}) = (x - A_x)^i (y& - A_y)^j (z - A_z)^k\\& \exp{ - \xi (\mathbf{r} - \mathbf{A})^2}\,,
    \end{split}
    \end{equation}
    which we separate into three independent Gaussians, 
    \begin{equation}
        \begin{split}
            G_{ijk} (\mathbf{r}, \xi, \mathbf{A}) = G_i(x, \xi, A_x) G_j (y, \xi,& A_y)\\&G_k(z, \xi, A_z)\ ,
        \end{split}
    \end{equation}
    where $\displaystyle G_i(x, \xi, A_x) = (x - A_x)^i \exp{-\xi(x - A_x)^2}.$ This separation will be instrumental in obtaining an analytical form for calculation of atomic orbital overlaps.
    
    It is useful to define Hermite Gaussian functions (see~\cite{GaussianBook} for more details), 
    \begin{equation}
        \Lambda_t(x, \xi, A_x) = \left(\pdv{}{A_x}\right)^t \exp{-\xi(x - A_x)^2}\ .
    \end{equation}
    These Hermite Gaussians will appear below and are related to Hermite polynomials $H_t(x)$ as
    \begin{equation}
        \Lambda_t (x, \xi, A_x) = \sqrt{\xi} H_t\left(\sqrt{\xi} (x - A_x)\right)\exp{-\xi(x - A_x)^2}.
    \end{equation}
    
    We now discuss the overlap between two orbitals, 
    \begin{equation}
    \begin{split}
        \Omega_{ij}(x, a, b, A_x, B_x) \equiv& G_i(x, a, A_x) G_j (x, b, B_x)\\
        =& x_A^i x_B^j \exp{-ax_A^2}\exp{-bx_B^2}\ ,
    \end{split}
    \end{equation}
    where $x_A\equiv x - A_x$. Now,
    \begin{equation}\label{eq:exp-axA2}
        \exp{-ax_A^2}\exp{-bx_B^2} = \exp{-qQ_x^2}\exp{-px_P^2}\ ,
    \end{equation}
    where
    \begin{equation}
        \begin{split}
            pP_x &= aA_x+bB_x\ ,\\
            Q_x &= A_x - B_x\ ,\\
            p &= a+b\ ,\\
            \text{and}\qquad q &= \frac{ab}{a+b}\ .
        \end{split}
    \end{equation}
    Note that the $x$-dependence in Eq.~\eqref{eq:exp-axA2} only comes from $\exp{-px_P^2}$, and so we can define the constant $K_{AB}\equiv\exp{-qQ_x^2}$. Thus we can write (see Equs.~49-53, 59, 60, and 70-75 in~\cite{GaussianBook})
    \begin{equation}
        \Omega_{ij} (x, a, b, A_x, B_x) = \sum_{t = 0}^{i+j} E_t^{ij} \Lambda_t (x, p, P_x)\ , 
    \end{equation}
    where the expansion coefficients have the recurrence relations,
    \begin{equation}
        \begin{split}
            E_0^{00} &= K_{AB}\ ,\\
            E_t^{i+1, j} &= \frac{1}{2p}E_{t-1}^{ij} - \frac{qQ_x}{a}E^{ij}_t + \left(t+1\right)E_{t+1}^{ij} ,\\
            E_t^{i+1, j} &= \frac{1}{2p}E_{t-1}^{ij} + \frac{qQ_x}{b}E^{ij}_t + \left(t+1\right)E_{t+1}^{ij} .
        \end{split}
    \end{equation}
    
    We can now finally calculate integrals of the form
    \begin{equation}
    \begin{split}
        &\mel{G_i (x, a, A_x)}{\exp{i k_x x_C}}{G_j(x, b, B_x)} \\
        &\qquad = \int_{-\infty}^\infty dx\ \Omega_{ij}(x, a, b, A_x, B_x) \exp{i k_x x_C}\\
        &\qquad =\sum_{t=0}^{i+j}E_t^{ij} \int_{-\infty}^\infty dx\ \Lambda_t(x, p, P_x) \exp{i k_x x_C}\\
        &\qquad =  \sum_{t=0}^{i+j}E_t^{ij} K_t^x,
    \end{split}
    \end{equation}
    where $\displaystyle K_t^x \equiv \int_{-\infty}^\infty dx\ \Lambda_t(x, p, P_x) \exp{i k_x x_C}.$ Expanding the Hermite Gaussian, we get
    \begin{equation}
        \begin{split}
            K_t^x =& \left(\pdv{P_x}\right)^t \int_{-\infty}^\infty dx\ \exp{ik_x x_C - px_P^2}\\
            =& \left(\pdv{P_x}\right)^t \exp{ik_x X_{PC}} \int_{-\infty}^\infty dx\ \exp{ik_x x_P - px_P^2} ,
        \end{split}
    \end{equation}
    with $X_{PC}\equiv P_x - C_x$. The integral term is now independent of $P_x$, and so the differential only applies to $\exp{ik_xX_{PC}}$, giving
    \begin{equation}
        K_t^x = \sqrt{\frac{\pi}{p}}\exp{ik_xX_{PC} - \frac{k_x^2}{4p}}\ \left(ik_x\right)^t .
    \end{equation}
    This gives us the analytical solution to the matrix element between two primitive gaussians, 
    \begin{equation}
        \begin{split}
            &\mel{G_i (x, a, A_x)}{\exp{i k_x x_C}}{G_j(x, b, B_x)} \\
            &\qquad = \sqrt{\frac{\pi}{p}}\exp{ik_xX_{PC} - \frac{k_x^2}{4p}}\ \sum_{t = 0}^{i+j} E_t^{ij} \left(ik_x\right)^t\ ,
        \end{split}
    \end{equation}
    which can be plugged into Eq.~\eqref{eq:mat_elems}, 
    \begin{equation}
        \begin{split}
            &f_{\left[j\mathbf{k}^\prime,i\mathbf{k}\right]}(\mathbf{q}) = \\
            &\qquad\sum_\mathbf{R}e^{i\mathbf{k}^\prime\cdot\mathbf{R}} \sum_{\alpha}\sum_\beta C^\dagger_{\beta i}(\mathbf{k})C_{j_\alpha}(\mathbf{k}^\prime)\sum_{\mu \in \beta}\sum_{\nu \in \alpha} N_\mu c_\mu N_\nu c_\nu\\
            &\qquad\times\mel{G_\mu^x (x, \xi_\mu, A_{\mu x})}{\exp{i q_x x}}{G_\nu^x (x, \xi_\nu, A_{\nu x})}\\
            &\qquad\times\mel{G_\mu^y (y, \xi_\mu, A_{\mu y})}{\exp{i q_y y}}{G_\nu^y (y, \xi_\nu, A_{\nu y})}\\
            &\qquad\times\mel{G_\mu^z (z, \xi_\mu, A_{\mu z})}{\exp{i q_z z}}{G_\nu^z (z, \xi_\nu, A_{\nu z})}\ .
        \end{split}
    \end{equation}
    
\section{Derivation of Scattering Rate Formulae}\label{app:derivation}
    
    In this section, we will briefly review derivation of the scattering rate formulae, mostly following Appendix A of~\cite{Essig:2015cda}. If a DM particle scatters with an electron in a stationary bound state, such as in a crystal, it can excite the electron from some initial energy $E_{e,1}$ to some final energy $E_{e,2}$ by transferring four-momentum $(E_e, \Vec{q})$. We describe the derivation in the context of field theory, treating the electron as being bound in a static background potential -- in other words, treating it non-relativistically during the interaction. This is a valid approximation because the momentum transfers are $q \sim \order{\text{several keV}} \ll m_e$. 
    
    \subsection{General formula for DM induced transitions}
        The cross section for free $2\longrightarrow2$ scattering is given by
        \begin{equation}
            \begin{split}
                \sigma v_\mathrm{free} =&\ \frac{1}{4E_\chi^\prime E_e^\prime}\int\frac{d^3q}{(2\pi)^3}\frac{d^3k^\prime}{(2\pi)^3}\frac{1}{4E_\chi E_e}\abs{\overline{\mathcal{M}_\mathrm{free}(\mathbf{q})}}^2\\
                &\ \ \ \times(2\pi)^4\delta(E_i - E_f)\delta^3(\mathbf{k-q-k}^\prime)\ ,
            \end{split}
        \end{equation}
        where $\mathcal{M}_\mathrm{free}$ is the field-theory matrix element and $\abs{\overline{\mathcal{M}_\mathrm{free}}}^2$ is its absolute squared averaged over initial spins and summed over final spins.
        
        For bound electron initial and final states, say $\psi_1$ and $\psi_2$, respectively, the cross section is modified as
        \begin{equation}
            \begin{split}
                V (2\pi)^3 \delta^3(\mathbf{k - q  - k}^\prime) &\abs{\mathcal{M}_\mathrm{free}}^2 \longrightarrow\\ & V^2 \abs{\mathcal{M}_\mathrm{free}}^2 \abs{f_{1\rightarrow2} (\mathbf{q})}^2\ ,
            \end{split}
        \end{equation}
        where 
        \begin{equation}
            f_{1\rightarrow2} (\mathbf{q}) \equiv \int d^3x\ \psi_2^*(\mathbf{x})e^{i\mathbf{q\cdot x}}\psi_1(\mathbf{x})\ .
        \end{equation}
        Moreover, because there is only one electron final state being considered, we can make the replacement $\displaystyle V\int \frac{d^3k^\prime}{(2\pi)^3} \longrightarrow 1$. 
        
        Combining these observations, we obtain
        \begin{equation}
            \begin{split}
                \sigma v_\mathrm{1\rightarrow2} = \frac{1}{4E_\chi^\prime E_e^\prime}\int\frac{d^3q}{(2\pi)^3}&\frac{1}{4E_\chi E_e}2\pi\delta(E_i - E_f)\times\\
                &\abs{\overline{\mathcal{M}_\mathrm{free}(\mathbf{q})}}^2\abs{f_{1\rightarrow2} (\mathbf{q})}^2\ .
            \end{split}
        \end{equation}
        For non-relativistic scattering, 
        \begin{equation}
            \begin{split}
                E_i =&\ m_\chi + m_e + \frac{1}{2}m_\chi v^2 + E_{e,1}, \text{ and}\\
                E_f =&\ m_\chi + m_e + \frac{\abs{m_\chi\mathbf{v - q}}^2}{2m_\chi} + E_{e,2}\ .
            \end{split}
        \end{equation}
        Moreover, we can parametrize 
        \begin{equation*}
            \abs{\overline{\mathcal{M}_\mathrm{free}(\mathbf{q})}}^2 = \abs{\overline{\mathcal{M}_\mathrm{free}(\alpha m_e)}}^2\times \abs{F_\chi (\mathbf{q})}^2\ ,
        \end{equation*}
        \begin{equation}
            \text{and }\Bar{\sigma}_e \equiv \frac{\mu_{\chi e}^2\abs{\overline{\mathcal{M}_\mathrm{free}(\alpha m_e)}}^2}{16\pi m_\chi^2 m_e^2}\ ,
        \end{equation}
        so the cross section simplifies to 
        \begin{equation}\label{eq:cross_section}
            \begin{split}
                \sigma v_\mathrm{1\rightarrow2} = \frac{\Bar{\sigma}_e}{\mu_{\chi e}^2} \int \frac{d^3q}{4\pi} & \delta\left(E_e + \frac{q^2}{2m_\chi} - \mathbf{q\cdot v}\right)\times\\
                &\abs{F_\chi (\mathbf{q})}^2\abs{f_{1\rightarrow2} (\mathbf{q})}^2 \ .
            \end{split}
        \end{equation}
    
    \subsection{Average rate in a DM halo}
        The rate of the specific transitions induced by DM hitting a target electron is then
        \begin{equation}
            R_{1\rightarrow2} = n_\chi \int d^3 v g_\chi(\mathbf{v}) \sigma v_{1\rightarrow2}\ ,
        \end{equation}
        where $n_\chi$ and $g_\chi(\mathbf{v})$ are the DM number density and velocity distribution, respectively. In this work, we use the parameters recommended by~\cite{Baxter2021}. 
        
        Note that the velocity distribution of DM in the standard halo model implies that the speed of the DM wind we observe must follow $v_\chi < v_\mathrm{Escape} + v_\odot + v_\oplus = v_\mathrm{Escape} + v_\mathrm{Earth}$, where $v_\mathrm{Escape}$ is the escape velocity at the Sun's location in the galactic gravitational potential well, $v_\odot$ is the sun's galactocentric speed, $v_\oplus$ is the earth's heliocentric speed, and $v_\mathrm{Earth}$ is the Earth's galactocentric speed.
        
        In this paper, we assume both DM velocity distribution and electron wavefunctions to be spherically symmetric, which is not true in general. We then use the integral over DM velocity to eliminate the $\delta$--function in Eq.~\eqref{eq:cross_section}, obtaining 
        \begin{equation}
            \begin{split}
                R_{1\rightarrow2} = \frac{n_\chi \Bar{\sigma}_e}{\mu_{\chi e}^2}&\int\frac{d^3 q}{4\pi}\int\frac{v^2 dv d\phi_v}{qv}g_\chi(\mathbf{v})\times \\
                &\Theta\left(v - v_\mathrm{min}(q, E_e)\right)\abs{F_\chi (\mathbf{q})}^2 \abs{f_{1\rightarrow2}(\mathbf{q})}^2\ .
            \end{split}
        \end{equation}
        Here $v_\mathrm{min}$ is the minimum velocity of the DM particle required for an energy-momentum transfer of $(E_e, \mathbf{q})$ to be feasible, 
        \begin{equation}
            v_\mathrm{min}(q, E_e) = \frac{E_e}{q} + \frac{q}{2m_\chi}\ .
        \end{equation}
        We define
        \begin{equation}
            \eta\left(v_\mathrm{min}(q, E_e)\right) \equiv \int \frac{d^3v}{v}g_\chi(\mathbf{v})\Theta\left(v - v_\mathrm{min}(q, E_e)\right)\ ,
        \end{equation}
        and obtain
        \begin{equation}\label{eq:rate_in_halo}
            \begin{split}
                R_{1\rightarrow2} = \frac{n_\chi \Bar{\sigma}_e}{8\pi\mu_{\chi e}^2}\int d^3 q \frac{1}{q} &\eta\left(v_\mathrm{min}\left(q, E_e\right)\right)\times\\
                &\abs{F_\chi\left(\mathbf{q}\right)}^2\abs{f_{1\rightarrow2}(\mathbf{q})}^2\ .
            \end{split}
        \end{equation}
        
    \subsection{Excitation rates in crystals}
        So far, we have not discussed the initial and final states of the electron being acted upon. In a crystal system, an electron may transition from an occupied orbital (core or valence) to an unoccupied state (conduction or free). Since we treat both core and valence shells equivalently, we simply call these occupied orbitals. Using the setup discussed in \S\ref{subsec:2.1}, we describe now the transition form factors, $f_{1\rightarrow2}(\mathbf{q})$.
        
        The electron is excited from an occupied state $\ket{\psi_{i\mathbf{k}}}$ to a conduction state $\ket{\psi_{i^\prime\mathbf{k}^\prime}}$, and so $f_{1\rightarrow2}(\mathbf{q}) \longrightarrow f_{i\mathbf{k}\rightarrow i^\prime\mathbf{k}^\prime}$, with 
        \begin{equation}
            \begin{split}
                f_{i\mathbf{k}\rightarrow i^\prime\mathbf{k}^\prime} =&\ \mel{\psi_{i^\prime\mathbf{k}^\prime}}{\exp{i\mathbf{q\cdot r}}}{\psi_{i\mathbf{k}}}\\
                =&\ \frac{1}{N_\mathrm{cell}} C_{i^\prime \beta}^\dagger(\mathbf{k^\prime})\mel{\phi_{\beta \mathbf{k}^\prime}}{e^{i\mathbf{q\cdot r}}}{\phi_{\alpha \mathbf{k}}}C_{\alpha i}(\mathbf{k})\ ,
            \end{split}
        \end{equation}
        where we have maintained the PySCF normalization, and Einstein summation over $\alpha$ and $\beta$ indices is implied. We shall invoke the orthogonality relation
        \begin{equation}
            \frac{V_\mathrm{cell}}{(2\pi)^3}\sum_\mathbf{R}e^{i\mathbf{q\cdot R}} = \sum_\mathbf{G}\delta^3\left(\mathbf{q - G}\right)\ .
        \end{equation}
        The matrix element is then given by 
        \begin{widetext}
            \begin{equation}
                \begin{split}
                    f_{i\mathbf{k}\rightarrow i^\prime\mathbf{k}^\prime} (\mathbf{q}) =&\ \frac{1}{N_\mathrm{cell}}C_{i^\prime \beta}^\dagger(\mathbf{k^\prime}) \sum_{\mathbf{R}} \sum_{\mathbf{R^\prime}}e^{-i\mathbf{k^\prime\cdot R^\prime}} e^{i\mathbf{k\cdot R}}\int d^3r\ \Tilde{G}_\beta^*(\mathbf{r - R^\prime})e^{i\mathbf{q\cdot r}}\Tilde{G}_\alpha(\mathbf{r - R})\ C_{\alpha i}(\mathbf{k})\\
                    =&\ \frac{1}{N_\mathrm{cell}}C_{i^\prime \beta}^\dagger(\mathbf{k^\prime}) \sum_{\mathbf{R}}e^{i\mathbf{\left(k+q-k^\prime\right)\cdot R}} \sum_{\mathbf{R^\prime}}e^{-i\mathbf{k^\prime\cdot R^\prime}}\int d^3 r\ \Tilde{G}_\beta^* (\mathbf{r - R^\prime})e^{i\mathbf{q\cdot r}}\Tilde{G}_\alpha (\mathbf{r})\ C_{\alpha i}(\mathbf{k})\\
                    =&\ \frac{(2\pi)^3}{V}\sum_{\mathbf{G}}\delta^3\left(\mathbf{k + q - k^\prime - G}\right)\sum_\mathbf{R}e^{-i\mathbf{k^\prime\cdot R}}\int d^3r\ C^\dagger_{i^\prime\beta}(\mathbf{k}^\prime) \Tilde{G}_\beta^*(\mathbf{r - R}) e^{i\mathbf{q\cdot r}} \Tilde{G}_\alpha(\mathbf{r})C_{\alpha i}(\mathbf{k})\\
                    =&\ \frac{(2\pi)^3}{V} f_{\left[i^\prime\mathbf{k}^\prime,i\mathbf{k}\right]}(\mathbf{q}) \sum_{\mathbf{G}} \delta^3\left(\mathbf{k + q - k^\prime - G}\right)\ ,
                \end{split}
            \end{equation}
        \end{widetext}
        where we have used the definition of $f_{\left[i^\prime\mathbf{k}^\prime,i\mathbf{k}\right]}(\mathbf{q})$ from Eq.~\eqref{eq:mat_elems}. We can now plug this into Eq.~\eqref{eq:rate_in_halo} to obtain
        \begin{equation}
            \begin{split}
                R_{i\mathbf{k}\rightarrow i^\prime \mathbf{k^\prime}} = &\frac{\pi^2 n_\chi \Bar{\sigma}_e}{V\mu_{\chi e}^2}\sum_\mathbf{G}\frac{1}{q}\eta\left(v_\mathrm{min}(q, E_{i^\prime\mathbf{k^\prime}} - E_{i\mathbf{k}})\right)\\
                &\left.\times\abs{F_\chi(\mathbf{q})}^2\abs{f_{i\mathbf{k}\rightarrow i^\prime\mathbf{k}^\prime} (\mathbf{q})}^2\right|_{\mathbf{q = k^\prime + G - k}}\ .
            \end{split}
        \end{equation}
        
        To calculate the total event rate, we must sum over occupied orbitals $i$ and unoccupied orbitals $i^\prime$, and integrate over both $\mathbf{k}$ and $\mathbf{k^\prime}$. Moreover, we must also consider the spin of the electrons in the occupied bands, giving
        \begin{equation}
            R_\mathrm{crystal} = 2\sum_i^\mathrm{occ}\sum_{i^\prime}^\mathrm{unocc}\int_\mathrm{BZ}\frac{Vd^3k}{(2\pi)^3}\int_\mathrm{BZ}\frac{Vd^3k^\prime}{(2\pi)^3}R_{i\mathbf{k}\rightarrow i^\prime \mathbf{k^\prime}}\ .
        \end{equation}
        Expanding this, and inserting the relevant $\delta-$distributions in $\mathbf{q}$ and $E_e$,
        \begin{widetext}
            \begin{equation}
                \begin{split}
                    R_\mathrm{crystal} = \frac{2\pi^2n_\chi\Bar{\sigma}_e}{\mu_{\chi e}}V\int_{-\infty}^\infty d\ln E_e\ E_e\int d^3q \frac{1}{q} \eta(v_\mathrm{min}(q, E_e))\abs{F_\chi(\mathbf{q})}^2\sum_{ii^\prime}&\int_\mathrm{BZ}\frac{d^3k\ d^3k^\prime}{(2\pi)^6}\delta(E_e - (E_{i^\prime\mathbf{k^\prime}} - E_{i\mathbf{k}}))\times\\
                    &f_{\left[i^\prime\mathbf{k}^\prime,i\mathbf{k}\right]}(\mathbf{q}) \sum_{\mathbf{G}} \delta^3\left(\mathbf{k + q - k^\prime - G}\right)\ .
                \end{split}
            \end{equation}
        \end{widetext}
        
        To simplify this form, we take two steps. First, we define $\mathbf{U\equiv k^\prime+G - k},$ and so $\delta^3\left(\mathbf{k + q - k^\prime - G}\right) = \delta^3(\mathbf{q - U})$, which can be further expanded as
        \begin{equation}
            \begin{split}
                \delta^3(\mathbf{q - U}) = \frac{1}{q^2 \sin{\theta_q}}\delta(q-U)\delta(\theta_q - \theta_U)\delta(\phi_q - \phi_U)\ .
            \end{split}
        \end{equation}
        Here $\theta_U$ and $\phi_U$ have the usual definitions as the inclination and azimuthal angles, respectively. From here, we can integrate over $\Omega_q$. Second, we differentiate the rate equation with respect to $\ln E_e$, and obtain Eqs.~\eqref{eq:theoryFF} and \eqref{eq:finRate},
        \begin{widetext}
            \begin{equation}
                \dv{R_\mathrm{crystal}}{\ln{E_e}} = n_\chi N_\mathrm{cell} \Bar{\sigma}_e \alpha \frac{m_e^2}{\mu_{\chi e}^2}\int d\ln{q}\ \frac{E_e}{q}\eta\left(v_\mathrm{min}\left(q, E_e\right)\right)\abs{F_\chi(q)}^2\abs{f_\mathrm{crystal}(q, E_e)}^2\ , \qquad\text{where}
            \end{equation}
            \begin{equation}
                \begin{split}
                    \left|f_{\rm crystal}(q, E_e)\right|^2 \equiv \frac{2\pi^2}{E_e}\frac{1}{\alpha m_e^2 V_\mathrm{cell}}\sum_{ii^\prime}\int_{\rm BZ}\frac{V_\mathrm{cell}d^3k}{(2\pi)^3}&\frac{V_\mathrm{cell}d^3k^\prime}{(2\pi)^3}E_e\delta\left(E_e - (E_{j\mathbf{k}^\prime} - E_{i\mathbf{k}})\right) \times\\
                    &\left.\sum_\mathbf{G^\prime}q\delta\left(q -\left|\mathbf{k^\prime+G^\prime - k}\right|\right)\left|f_{\left[i^\prime\mathbf{k}^\prime,i\mathbf{k}\right]}(\mathbf{q})\right|^2\right|_{\theta_q = \theta_U, \phi_q = \phi_U}\ .
                \end{split}
            \end{equation}
        \end{widetext}
\end{document}